\documentclass[twocolumn, amssymb, nobibnotes, aps, prd, nofootinbib, longbibliography]{revtex4-2}
\usepackage[T1]{fontenc}
\usepackage{amsmath}
\usepackage{amsfonts}
\usepackage{amssymb}
\usepackage{blindtext}
\usepackage{multirow}
\usepackage[colorlinks=true, linkcolor=blue, urlcolor=blue, citecolor=blue]{hyperref}
\usepackage{graphicx}
\usepackage[dvipsnames]{xcolor}
\usepackage{tikz}
\usetikzlibrary{shapes.geometric, arrows}

\tikzstyle{process} = [rectangle, 
minimum width=2cm, 
minimum height=1cm, 
text centered, 
text width=2cm, 
draw=black,]

\tikzstyle{decision} = [diamond, 
minimum width=3cm, 
minimum height=1cm, 
text centered, 
draw=black,]
\tikzstyle{arrow} = [thick,->,>=stealth]

\newcommand{\dcc}{P2200371-v3} 
\newcommand{\tds}{ET-0085B-23} 
\newcommand{\tdsurl}{https://apps.et-gw.eu/tds/?content=3&r=18220} 
\newcommand{\orc}{\includegraphics[height=\fontcharht\font`A]{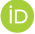}}
\newcommand{\orcid}[1]{\href{https://orcid.org/#1}{\orc}}

\newcommand{\Fstat}{$\mathcal{F}$-statistic\xspace}
\newcommand{\Fstatmax}{\ensuremath{2\mathcal{F}_{\textrm{max}}}\xspace}
\newcommand{\win}{\varpi}
\newcommand{\winrect}{\win_\textrm{r}}
\newcommand{\winexp}{\win_\textrm{e}}
\newcommand{\Dop}{\lambda} 
\newcommand{\Amp}{\mathcal{A}} 
\newcommand{\TP}{\mathcal{T}} 
\newcommand{\Btrans}{B_{\mathrm{tS}/\mathrm{G}}} 
\newcommand{\fgw}{f} 
\newcommand{\fdot}{{\dot{\fgw}}} 
\newcommand{\fddot}{\ddot{\fgw}} 
\newcommand{\fdddot}{\dddot{\fgw}} 
\newcommand{\rhocnn}{\rho_{\textrm{CNN}}}

\begin{document}
\title{Convolutional neural network search for long-duration transient gravitational waves from glitching pulsars}
\author{Luana M. Modafferi\,\orcid{0000-0002-3422-6986}} 
\email{luana.modafferi@ligo.org}
\affiliation{Departament de F\'isica, Institut d'Aplicacions Computacionals i de
Codi Comunitari (IAC3), Universitat de les Illes Balears, and Institut d'Estudis
Espacials de Catalunya (IEEC), Carretera de Valldemossa km 7.5, E-07122 Palma,
Spain}
\author{Rodrigo Tenorio\,\orcid{0000-0002-3582-2587}}
\email{rodrigo.tenorio@ligo.org}
\affiliation{Departament de F\'isica, Institut d'Aplicacions Computacionals i de
Codi Comunitari (IAC3), Universitat de les Illes Balears, and Institut d'Estudis
Espacials de Catalunya (IEEC), Carretera de Valldemossa km 7.5, E-07122 Palma,
Spain}
\author{David Keitel\,\orcid{0000-0002-2824-626X}} 
\email{david.keitel@ligo.org}
\affiliation{Departament de F\'isica, Institut d'Aplicacions Computacionals i de
Codi Comunitari (IAC3), Universitat de les Illes Balears, and Institut d'Estudis
Espacials de Catalunya (IEEC), Carretera de Valldemossa km 7.5, E-07122 Palma,
Spain}

\date{
dated 2023-03-29
-- \href{https://dcc.ligo.org/\dcc}{LIGO-\dcc} -- \href{\tdsurl}{\tds}
}

\begin{abstract}
Machine learning can be a powerful tool to discover new signal types in
    astronomical data. We here apply it to search for long-duration transient gravitational
waves triggered by pulsar glitches, which could yield physical
insight into the mostly unknown depths of the pulsar. Current methods to search for such signals
rely on matched filtering and a brute-force grid search over possible
signal durations, which is sensitive but can become very computationally
    expensive. We develop a method to search for post-glitch signals
on combining matched filtering with convolutional neural networks, which reaches similar
    sensitivities to the standard method at false-alarm probabilities relevant for practical searches,
while being significantly faster.
We specialize to the Vela glitch during the LIGO--Virgo
    O2 run, and set upper limits on the gravitational-wave strain amplitude from the data of the two LIGO
    detectors for both constant-amplitude and exponentially decaying signals.
\end{abstract}

\maketitle

\section{Introduction}
\label{sec:introduction}
Gravitational wave (GW) detectors are sensitive to a variety of different sources.
Spinning neutron stars (NSs)
with non-axisymmetric deformations, in particular, are promising
GW sources \cite{Glampedakis:2017nqy},
e.g. for long-lasting quasi-monochromatic
signals (continuous waves, CWs) \cite{Haskell:2021ljd,Riles:2022wwz}.
The expected amplitudes of CWs are several orders of magnitude smaller
than the signals already detected from compact binary coalescences (CBCs)
\cite{LIGOScientific:2021djp}, requiring long integration times, and have not
been discovered yet by current LIGO--Virgo--KAGRA detectors
\cite{aLIGo,VIRGO_detect, KAGRA}.

Pulsars are rotating NSs that also emit electromagnetic (EM) beams.
Of these, some show glitches \cite{main_1},
i.e. rare anomalies in their otherwise very stable
frequency evolution.
Glitches typically feature a sudden spin-up, often followed by a relaxation phase,
and are one of the few circumstances in which
we may examine the inside of a NS and the characteristics of matter at
supernuclear density \cite{haskell_2017}. They could also excite
non-axisymmetric perturbations, hence triggering GWs \cite{Haskell:2015jra}. These
signals could be ``burst-like'', with dampening timescales of milliseconds
\cite{10.1046/j.1365-8711.1998.01840.x}, and also long-duration transient CW-like signals (tCWs)
with timescales from hours to months \cite{Prix:2011qv}. The detection of GWs
from a pulsar glitch could yield more understanding of the NS
equation of state \cite{vanEysden:2008pd}.

Detection prospects for tCWs were studied in Ref.~\cite{10.1093/mnras/stac3665},
finding that upcoming LVK observing runs will offer the first chances to detect them
or at least to obtain upper limits that can physically constrain glitch models.
Third generation GW detectors such as Einstein Telescope \cite{Punturo_2010} and Cosmic Explorer \cite{Reitze:2019iox} will be able to probe deeper into the population of known glitchers.
The Vela pulsar (J0835$-$4510) is a particularly strong candidate, because of its
large and frequent glitches, also characterized by long recovery times up to several hundred days.

Recent works have performed unmodelled searches for burst-like
transients \cite{KAGRA:2021tnv,PhysRevD.106.103037}
and modelled searches for long-duration tCWs
\cite{Keitel:2019zhb,LIGOScientific:2021quq,Modafferi:2021hlm}. In this work we focus on
tCW searches, which are so far based on matched filtering
of the data against a signal model.
Typically, a template bank is constructed to cover the possible range of signal parameters.
However these model-dependent searches are usually limited in the volume of
parameter space that can be covered, due to high computational cost, as we detail in
Sec.~\ref{sec:search}.

A possible way out of computational bottlenecks in GW data analysis is
deep learning (DL), a subfield of machine learning
which consists of processing data with deep neural networks. DL has been
gaining ground in different scientific fields and in particular in gravitational
astronomy, as it offers powerful novel methods to search for or classify signals
while decreasing computational cost. Ref.~\cite{Cuoco:2020ogp} offers a review of
GW applications, spanning from data quality
studies and detector characterization, to waveform modeling and
searches using different DL model architectures. In this work we provide a DL
algorithm based on a Convolutional Neural Network (CNN) as a complementary tool
to matched filtering in the search for tCWs from glitching pulsars.

The
paper is structured as follows. In Sec.~\ref{sec:search} we set up the detection
problem with a specific real-world test case
(searching for tCWs after the 2016 Vela pulsar glitch),
starting from matched filtering,
and we motivate how DL can help with its limitations.
In Sec.~\ref{sec:training} we introduce our CNN architecture and training
strategy and in Sec.~\ref{sec:evaluation_method} we explain our evaluation method and testing sets.
In Sec.~\ref{sec:results} we test the CNN on simulated and real data,
and in Sec.~\ref{sec:parameter_estimation} we consider extensions to parameter estimation.
We then present observational upper limits on the strain amplitude of tCWs
after the 2016 Vela glitch in Sec.~\ref{sec:upper_limits},
comparing with those previously obtained in Ref.~\cite{Keitel:2019zhb}.
Finally we state conclusions in Sec.~\ref{sec:conclusions}.

\section{Problem statement and a real-world test case}
\label{sec:search}
The standard method to search for tCWs is based on matched filtering and was
first introduced in Ref.~\cite{Prix:2011qv}.
So far it has been applied in Ref.~\cite{Keitel:2019zhb,LIGOScientific:2021quq}
to search for signals from various known glitching pulsars
in data from the second and third LIGO--Virgo observing runs (O2 and O3).
We briefly summarize the method here and provide more details in Appendix
\ref{sec:fstat_derivation}. Considerations for the practical setup of such searches
based on the ephemerides of known pulsars are
provided in Ref.~\cite{Modafferi:2021hlm}. The setup of this detection method is very
similar to that of narrowband CW searches \cite{LIGOScientific:2021quq,2017PhRvD..96l2006A,O2Narrowband},
but for tCWs one must also take into
account additional parameters describing the temporal evolution of the signal,
e.g. its start time and duration. 

The pulsar population presents a large variety of glitching
behaviors~\cite{main_1}, and there are several models explaining this
phenomenology~\cite{Haskell:2015jra,haskell_2017}. Similarly, GW signals could
derive from different physical mechanisms, e.g. Ekman flows
\cite{models_1,models_2,Singh:2016ilt} or Rossby \textit{r-}modes
\cite{Andersson,Santiago-Prieto:2012qwb}, varying from glitch to glitch.
Depending on the model, it may be reasonable to expect tCWs with durations
of the order of the timescales of the post-glitch recovery,
i.e. the time needed for the frequency to return to its secular value \cite{Yim:2020trr}. In general,
searches for tCWs must account for this model uncertainty by covering wider ranges of possible
parameters, including all possible combinations of the transient parameters and
sufficiently wide frequency evolution template banks, of sizes up to
tens of millions per target.

\subsection{Current method for detecting tCWs}

In matched filtering the first step is to define the signal model. 
We ignore the specific physical process behind the generation of the signal, and
define a generic tCW model following Ref.~\cite{Prix:2011qv}, by generalizing the standard
CW model.

First we introduce 
\begin{equation}
    h_{\textrm{CW}}(t;\Dop,\Amp) = \sum_{\mu=0}^{3} \Amp^{\mu}h_{\mu}(t;\Dop), 
    \label{eq:hCW}
\end{equation}
where $h_{\mu}$ are four basis functions derived in Ref.~\cite{PhysRevD.58.063001},
and $\Dop, \Amp$ are the frequency evolution and amplitude parameters of the
signal, respectively\footnote{We follow the notation of \cite{Prix_2007, Prix:2011qv, Prix:2015cfs} and specialize to a single signal harmonic.}. The four amplitude parameters
depend on the CW amplitude $h_0$, inclination $\cos\iota$, polarization angle
$\psi$, and initial phase $\phi_0$, while the frequency evolution parameters
include the source position (right ascension $\alpha$ and declination $\delta$)
and the GW frequency $\fgw$ and spindowns (frequency derivatives)
$\fdot, \fddot, \fdddot, \dots$
For GW emission driven by a mass quadrupole, the GW frequency and spindown
values are twice the rotational values.

From Eq.~\eqref{eq:hCW}, tCWs can be obtained by multiplying with a window function $\win(t;\TP)$
dependent on time and an additional set of transient parameters $\TP$:
\begin{equation}
    h(t; \Dop, \Amp, \TP) = \win(t;\TP) h_{\textrm{CW}}(t;\Dop, \Amp).
    \label{eq:tCW_model}
\end{equation}
For instance, $\TP$ can be the start time $t^0$ and the duration $\tau$ of the tCW signal.
Two simple choices of window functions are the rectangular window, for signals of constant amplitude truncated in time,
\begin{equation}
    \winrect(t;t^0,\tau) = 
        \begin{cases}
            1 & \text{if $t\in [t^0, t^0+\tau]$}\\
            0 & \text{otherwise}
        \end{cases},
    \label{eq:winrect}
\end{equation}
and the exponentially decaying window
\begin{equation}
    \winexp(t;t^0, \tau) = 
        \begin{cases}
            e^{-(t-t^0)/\tau} & \text{if $t \in [t^0, t^0 +3\tau]$}\\
            0 & \text{otherwise}
        \end{cases},
    \label{eq:winexp}
\end{equation}
where for $\winexp$ the truncation at $3\tau$ was chosen by Ref.~\cite{Prix:2011qv}
as the point where the amplitude has decreased by more than 95\%. 

While the rectangular window follows the exact evolution of a standard CW and
solely cuts it in time, the exponentially decaying window modifies the time
evolution of the amplitude $h_0$. In general, one can define an arbitrary window
function depending on different transient parameters $\TP$, but for simplicity
we only consider and compare the two choices above.

Once the signal model is
defined, one can then proceed as in Ref.~\cite{Prix:2011qv} with a noise-versus-signal
hypothesis test, assuming Gaussian noise for the noise hypothesis and adding
Eq.~\eqref{eq:tCW_model} for the signal hypothesis. The standard procedure
consists of maximizing the likelihood ratio between the hypotheses over the
amplitude parameters $\Amp$, obtaining what is known as the \Fstat \cite{PhysRevD.58.063001,
Prix:2009tq}.

In general, for both CWs and tCWs,
the \Fstat over a given stretch of data is calculated from the
antenna pattern matrix $\mathcal{M}_{\mu\nu}$ and the projections $x_{\mu}$ of
the data onto the basis functions $h_{\mu}$. The implementation in \texttt{LALSuite}
\cite{LALSuite,Prix:2015cfs} splits the data $x^X(t)$ of detector $X$ into short Fourier transforms (SFTs) \cite{sfts_2022},
i.e. the Fourier Transforms of time segments of
duration $T_{\textrm{SFT}}$. Then the \Fstat is computed from a set of per-SFT
quantities numerically of order $\mathcal{O}(1)$, called \Fstat \textit{atoms}.
Coherently combining $N_{\textrm{det}}$ detectors,
and suppressing the dependence on $\lambda$ for the rest of this section,
one can write:
\begin{equation}
    \begin{aligned}
        x_{\mu,\alpha} = 2\sum_{X=1}^{N_{\textrm{det}}}S^{-1}_{X\alpha} \int_{t_{\alpha}}^{t_{\alpha}+T_{\textrm{SFT}}} x^X(t)h_{\mu}^X(t)dt,\\
        \mathcal{M}_{\mu\nu,\alpha} = 2\sum_{X=1}^{N_{\textrm{det}}}S^{-1}_{X\alpha} \int_{t_{\alpha}}^{t_{\alpha}+T_{\textrm{SFT}}} h^X_{\mu}(t)h_{\nu}^X(t)dt,
    \end{aligned}  
    \label{eq:def_atoms}  
\end{equation}
where $S^{-1}_{X\alpha}$ is the single-sided noise power spectral density (PSD)
for a detector $X$ at the template frequency at the start time $t_{\alpha}$ of
the SFT $\alpha$. More specifically, in the following, with the word ``atoms''
we refer to a set of closely related per-SFT quantities as used in the
\texttt{LALSuite} code, which also include noise weighting.
We define these explicitly in Appendix \ref{sec:fstat_derivation},
following Ref.~\cite{Prix:2015cfs}.

To compute the (transient) \Fstat over a window $\win(t^0,\tau)$, one needs partially
summed quantities
\begin{align}
    x_{\mu}(t^0, \tau) &= \sum_{\alpha} \win(t_{\alpha};t^0,\tau)x_{\mu,\alpha},\\
    \mathcal{M}_{\mu\nu}(t^0, \tau) &= \sum_{\alpha} \win^2(t_{\alpha};t^0,\tau) \mathcal{M}_{\mu\nu,\alpha},
    \label{eq:projections}
\end{align}
where the sums go over the set of SFTs matching the window. 
From these,
\begin{equation}
    2 \mathcal{F}(t^0, \tau) = x_{\mu}(t^0, \tau)\mathcal{M}^{\mu\nu}(t^0, \tau)x_{\nu}(t^0, \tau),
    \label{eq:fstat}
\end{equation}
where $\mathcal{M}^{\mu\nu}$ is the inverse matrix of $\mathcal{M}_{\mu\nu}$.

Inserting the template into this equation defines the optimal signal-to-noise ratio (SNR)
\begin{equation}
 \label{eq:snr}
    \rho_{\textrm{opt}} = \sqrt{\mathcal{A}^{\mu} \mathcal{M}_{\mu \nu} \mathcal{A}^{\nu}},
\end{equation}
where we have suppressed the explicit dependence on the transient parameters. In
the absence of a signal, \mbox{$\rho_{\textrm{opt}}=0$}.
In later sections, we will regularly refer to simulated signals as ``injections''
(into real or simulated data)
and to the optimal SNR of the template used to generate an injection as $\rho_{\textrm{inj}}$.

To get a detection statistic from Eq.~\eqref{eq:fstat} that only depends on
$\Dop$, for the CW case, one just takes the total sums over the full
observation time. But for tCWs we need to handle the case of
unknown $t^0$ and $\tau$.
To obtain a detection statistic that depends only on $\Dop$, one can e.g. maximize over $\TP$
obtaining a statistic \Fstatmax, or marginalize over the same parameters obtaining the
transient Bayes factor $\Btrans$. It has been shown that $\Btrans$
is more sensitive than \Fstatmax in Gaussian noise \cite{Prix:2011qv} and also more robust on 
real data \cite{Tenorio:2021wad}.

\subsection{Computational cost limitations}

The method can be divided into three steps, at each $\Dop$: computing the \Fstat
atoms, computing Eq.~\eqref{eq:fstat} for all the different combinations of
transient parameters, and finally maximization/marginalization over the same
parameters to obtain an overall detection statistic.
Besides \texttt{LALSuite}, this has also been implemented in \texttt{PyFstat} \cite{Keitel:2021xeq},
which is a python package that wraps the corresponding \texttt{LALSuite} functions
and also adds a GPU implementation of the last two steps~\cite{Keitel:2018pxz}. 

The computing time varies for each step and can be estimated as discussed in
Ref.~\cite{Prix:2011qv,Keitel:2018pxz}. In particular, the most cost-intensive step
is typically the second one because of the large number of partial sums
that need to be taken corresponding to all different combinations of $\TP$.
Timing models and results for this step for both CPU and
GPU implementations can be found in Ref.~\cite{Keitel:2018pxz}.
As discussed there and in Ref.~\cite{Prix:2011qv},
this step also crucially depends on the window function:
calculations for exponential windows are much more expensive
not only because of the exponential functions themselves,
but also because partial sums cannot be efficiently reused
like in the rectangular case.
For realistic parameter ranges, searches with an exponential window model
can take orders of magnitude longer with respect to the
rectangular model in either implementation. Therefore, searches for
long-duration tCWs~\cite{Keitel:2019zhb,LIGOScientific:2021quq,Modafferi:2021hlm}
have only applied the simpler rectangular window, even though
models like that of Ref.~\cite{Yim:2020trr} assume post-glitch evolutions
following that of the EM signal, typically fitted by exponential functions.

DL models can avoid brute-force loops over many parameter combinations,
as once they are trained only a single forward-pass of the network is needed per input data instance.
Hence, they can be faster, and also have the potential to be more agnostic to signal models.
One could replace matched filtering completely by training a network on the full
detector data, thus allowing frequency evolutions different from the standard
spin-down model, but this would likely come with a loss in sensitivity, similarly to
excess power methods.

Instead, we apply CNNs with the \Fstat atoms as input data, 
which are an intermediate output of matched filtering, meaning we lock the
frequency evolution model but still allow flexibility in the amplitude evolution
of the GW and the potential for significant speed-up.
As we will demonstrate, with this approach one can reach
sensitivities similar to those of the standard detection statistics
at reduced computational cost. 

\subsection{Test case based on O2 Vela search}
\label{traditional_analysis}
We tune the setup of our training, search and comparisons to that from the first search for
long-duration tCWs \cite{Keitel:2019zhb}, using data
from the two LIGO detectors during the second
observing run (O2) of the LIGO--Virgo network \cite{LIGOScientific:2019lzm}.
Two priority
targets, Vela and the Crab, glitched during that observing period. No evidence
was found for tCW signals in the data, so 90\% upper limits on the GW
amplitude $h_0$ as a function of signal durations $\tau$ were reported. For a first
proof of concept we will only focus on the Vela (J0835$-$4510) analysis, since
its 2016 glitch is one of the largest that have been analyzed by GW searches,
with a relative jump in frequency of $\approx 1.43 \times 10^{-6}$. More glitches have
been analyzed during O3 \cite{LIGOScientific:2021quq}, of which two reached similar
strengths. But due to its combination of a large glitch and its proximity,
the 2016 Vela glitch was the most promising to date.

\begin{table}[]
    \begin{tabular}[t]{cc}
        \hline
        \hline
        \multicolumn{2}{c}{Source parameters} \\
        \hline
        \hline
        $d$ & $287$\,pc \\
        $\alpha$ & $2.2486$\,rad \\
        $\delta$ & $-0.7885$\,rad \\
        $f$ & $22.3722$\,Hz \\
        $\fdot$ & $-3.12\times 10^{-11}$\,Hz\,s$^{-1}$ \\
        $\fddot$ & $1.16\times 10^{-19}$\,Hz\,s$^{-2}$ \\
        $T_{\textrm{ref}}$ & $58000\,$MJD \\
        $T_{\textrm{gl}}$ & $57734.485$\,MJD \\
        \hline
        \hline
    \end{tabular}
    \quad
    \begin{tabular}[t]{cc}
        \hline
        \hline
        \multicolumn{2}{c}{Search parameters} \\
        \hline
        \hline
        $\Delta f$ & $0.1$\,Hz\\
        $\Delta \dot{f}$ & $1.01 \times 10^{-13}$\,Hz\,s$^{-1}$ \\
        $df$ &  $9.57 \times 10^{-8}$\,Hz \\
        $d\dot{f}$ & $9.15 \times 10^{-15}$\,Hz\,s$^{-1}$\\
        $t^0_{\textrm{min}}$ & $T_{\textrm{gl}} - 0.5$\,day\\
        $t^0_{\textrm{max}}$ & $T_{\textrm{gl}} + 0.5$\,day\\
        $\tau_{\textrm{min}}$ & $T_{\textrm{gl}}-\frac{1\,{\textrm{day}}}{2}+ 2T_{\textrm{SFT}}$\\
        $\tau_{\textrm{max}}$ & $120\,\textrm{days} - 2T_{\textrm{SFT}} $\\
        \hline
        \hline
    \end{tabular}
    \caption{Ephemerides and search parameters for Vela and its glitch during the O2 run. The distance
    from the source is indicated with $d$ and its sky coordinates are $\alpha$
    (right ascension) and $\delta$ (declination). The frequency and spindown
    values for the source parameters are the GW values and are all referenced to $T_{\textrm{ref}}$. 
    In the search parameters, we indicate with $\Delta$ the search band
    centered around the GW frequency (or spindown), and with $df, d\dot{f}$ the
    resolutions in frequency and spindown of the search. The search in second
    order spindown is fixed to the nominal GW value. For the transient
    parameters, we list the minimum and maximum values of start times $t^0$ and
    $\tau$ we search over, with resolutions matching $T_{\textrm{SFT}} = 1800$\,s.} 
    \label{tab:ephem}
    \end{table}

Following Ref.~\cite{Keitel:2019zhb}, we construct our test case as a narrow-band search
in frequency and a single spindown parameter,
centred on the values inferred from the timing observations of Ref.~\cite{2018Natur.556..219P},
and directed at Vela's sky position.
We search for tCWs with start times $t_0$ within half a day of the glitch time
(December 12th 2016 at 11:38:23.188 UTC), and with durations $\tau$ up
to 120\,days.
All relevant source and search parameters are listed in Table~\ref{tab:ephem}.
Even when using simulated data,
we match their timestamps to the real observational segments
as given by Ref.~\cite{O2SFTs}, 
which show a notable 12-day gap and various smaller ones.
We study the effect of these gaps on the various detection statistics in
Appendix~\ref{sec:gaps}.
The total number of analyzed SFTs, each of duration $T_{\textrm{SFT}}=1800$\,s,
are 3782 for H1 and 3234 for L1.

\section{CNN architecture and training}
\label{sec:training}
\subsection{Input data format}
The performance of DL models depends crucially on the type and quality of their
input data. In the GW field, different options have been explored depending on
how heavily the data have been transformed~\cite{Cuoco:2020ogp}.
Directly using GW strain time series
would allow a flexible training in terms of both amplitude and
frequency evolution of potential signals.
The whitened detector strain has been used e.g. in
Ref.~ \cite{Gebhard:2019ldz,PhysRevLett.120.141103,Schafer:2021fea} for CBC signals.
For CW-like signals, different approaches have been investigated. Some
examples include Fourier transforms of blocks of data in the search for long-duration
transients from newly-born neutron stars \cite{2019PhRvD.100f2005M}. Also, 
for the search of CWs from spinning NSs,
time-frequency spectrograms and Viterbi maps \cite{2020PhRvD.102h3024B}
and power-based statistics \cite{2019PhRvD.100d4009D, Yamamoto:2020pus, Yamamoto:2022adl}
have been used.

As stated in Sec.~\ref{sec:search}, the most expensive step of the algorithm
from Ref.~\cite{Prix:2011qv}
for post-glitch transients is typically the calculation of the partial
sums of the \Fstat atoms over the transient parameters $\TP$. In comparison, the
cost of computing the \Fstat atoms (the main matched-filter stage) is marginal.
Therefore we decide to
use as input data to our network the \Fstat atoms, which contain all the
information needed to determine the presence of a signal in the data. The exact
expressions of the atoms are derived in Appendix~\ref{sec:fstat_derivation}.
When passing to the network, we do not perform any type of preprocessing to the
\Fstat atoms, since they are already normalized to order $\mathcal{O}(1)$
both for noise-only data and for signals within the range of SNRs considered here.

As we will demonstrate, this choice allows for a comparatively simple DL model
architecture and training setup to reach detection sensitivity close to that of
traditional detection statistics. However, it means that we keep to the specific
frequency evolution of the standard CW model, and the atoms still have to be
computed for each set of parameters $\Dop$.

By using \Fstat atoms as input to the network,
we can also easily implement a multi-detector configuration by
merging the single-detector atoms into a set of equi-spaced timestamps
spanning the total observing time:
the result will correspond to
the single-detector values at timestamps when only one detector is available,
the sum when multiple detectors are on,
and filled up with zeros as required.
The merged atoms span 4190 timestamps, spaced by 1800\,s as the original SFTs.
The summing of the multi-detector atoms allows us to implement a simpler
network in contrast to e.g. multi-channel or multi-input models and comes with
the gain in sensitivity from coherently combining the detectors.

\subsection{CNN architecture}
\label{sec:architecture}
\begin{figure*}
    \includegraphics[width=\textwidth]{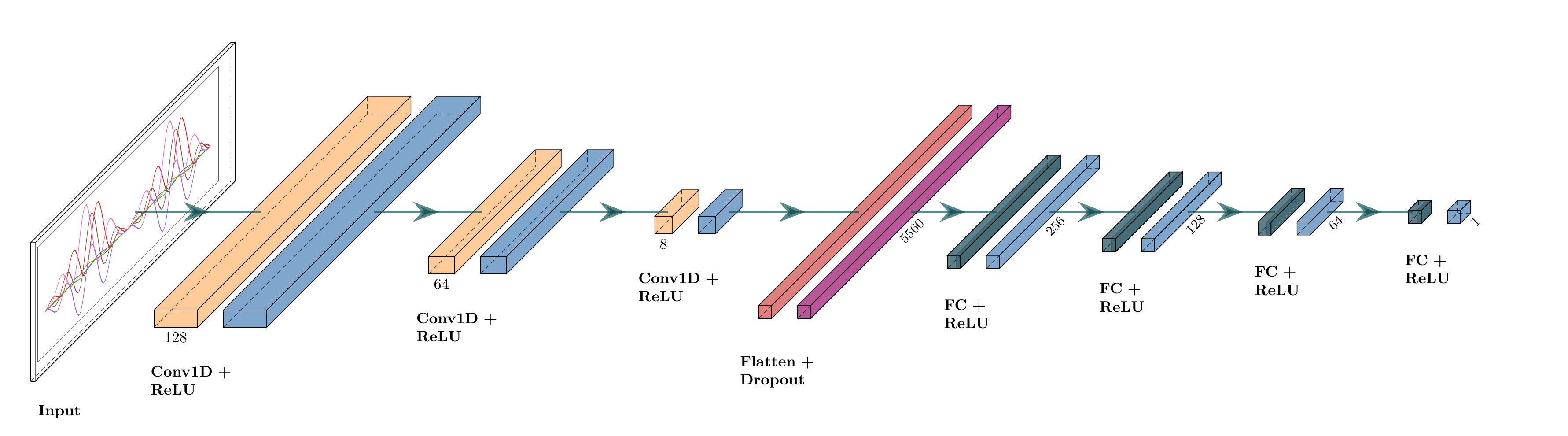}
    \caption{Architecture of the convolutional neural network (CNN) model,
    which as its input takes the \Fstat atoms. The CNN is made up of a stack of three
    convolutional layers (labelled Conv1D in figure) followed by four fully connected layers (FC) and the output
    layer. The convolutional and fully connected layers all use the ReLU activation function.
    The convolutional part and the fully connected part of the network
    are separated by a flattening layer, needed to transform the output feature
    maps to a digestible input for the dense layers, and a dropout layer.
    }
    \label{fig:cnn_architecture}
\end{figure*} 

CNNs are a type of artificial neural networks that use convolutional layers as
building blocks \cite{GoodBengCour16,cnn_indolia}. While in fully-connected
layers neurons are connected to all those of the previous layer, this would be
computationally unfeasible for a high-dimensional input such as an image, for
which a fully-connected layer would use a weight for each pixel. Convolutional
layers avoid this problem by exploiting the spatial structure and repeating
patterns in the data. To achieve this they convolve various independent filters
(also called convolution kernels) with the input. Each filter is a matrix of
weights to be trained, which slides along the input data, producing a feature map.
These maps are then transformed through a non-linear activation function.
This way, each filter learns to identify a pattern or characteristic of the data,
regardless of location,
which is preserved in the feature maps.
These maps are then stacked to yield the input for the next layer.
This enables the CNN to learn where features, or mixtures of features, are located in the data.

For this work we use a simple CNN,
made up of three one-dimensional convolutional layers.
This means that in each layer the filters slide along the time dimension (vertical axis)
and convolve all columns at once. 
We always set the filter size to 3 along the time dimension in each layer,
and to increasingly narrow down important features we use a decreasing number of 128, 64 and 8 independent filters per layer
and set the stride parameter to 3, 2 and 1 respectively.
The stride is the number of pixels the filter moves while sliding through the
input matrix, i.e. the bigger the stride, the smaller the output feature maps.
We do not implement pooling layers \cite{726791}, since we find that for our particular
architecture they are detrimental to the performance of the network.

The convolutional layers are then
followed by a flattening and dropout
layer \cite{dropout} and four fully-connected layers. 
Again we decrease the number of neurons from layer to layer (256, 128 and 64)
so that the number of trainable parameters
decreases when approaching the final output.
The last of the fully-connected layers
gives the output of the network. We choose this to be a single continuous,
positive value $\rho_{\textrm{CNN}}$
which we train to match the injected SNR from Eq.~\eqref{eq:snr}
for each sample in the signal training set, and to return 0 for pure-noise samples.
We use the Rectified Linear Unit (ReLU) activation function
\cite{2018arXiv180308375A} for all layers, including for the output,
which produces the desired range of values for an SNR-like quantity:
$\rho_{\textrm{CNN}}\in [0,+\infty)$.
The dropout layer, which helps to prevent the CNN from overfitting
\cite{dropout}, has a dropout rate of $0.33$.

The full network architecture is illustrated in
Fig.~\ref{fig:cnn_architecture} and has been implemented using the Keras
library \cite{chollet2015keras} based on the Tensorflow infrastructure \cite{tensorflow2015-whitepaper}.
For the various hyperparameters, we chose starting values from an exploratory
run with the optimization framework Optuna \cite{optuna_2019}, minimizing the
loss on the validation set using a sampler based on the Tree-structured Parzen
Estimator algorithm \cite{NIPS2011_86e8f7ab}, and further tuned them
manually.

\subsection{Training strategy}
\label{sec:training_strategy}
The CNN is trained by minimizing a loss function $L$ over a training set
containing both pure-noise samples and simulated signals.
We use the mean squared error loss:
\begin{equation}
    L = \frac{1}{n}\sum_{i=1}^n(\rho_{\textrm{inj}}-\rho_{\textrm{CNN}})^2 \,,
\end{equation}
where $\rho_{\textrm{inj}}$ is the injected SNR of the training signals,
corresponding to Eq.~\eqref{eq:snr}.
The training of our CNN's continuous output function can
therefore be considered a regression learning approach.

In general, we separate the training into two stages which differ with respect to the injection
set (both in the number of training samples and the SNR range of signal samples), optimizer
and number of epochs. A summary of the parameters used in each stage are shown
in Tab.~\ref{tab:curriculum} and will be explained in more detail below.

When training a network, the weights are updated by an optimizer through
gradient descent. There is a variety of well-established optimizers built to
e.g. prioritize speeding up convergence or improving generalization.
We use the Adam optimizer \cite{2014arXiv1412.6980K} in a
first training stage, i.e. for the first 200 epochs, and the Stochastic Gradient
Descent algorithm (SGD) \cite{2016arXiv160904747R} for the remaining 1000
epochs. We found that the Adam algorithm converges rapidly while the SGD
generalizes better, reducing the unwanted effect of overfitting
\cite{2017arXiv171207628S}.

We also use the curriculum learning (CL) training
strategy \cite{Bengio2009}. CL has been applied on GW data in previous works
\cite{PhysRevD.103.063011,PhysRevD.103.102003} and consists of training on datasets
of increasing difficulty. The difficulty criterion depends on the type
of data and problem, and for our case we choose the range of injected SNRs.
We use a simple two-stage CL
strategy, first training on an easier dataset with strong signals
(high SNR) and then adding a more difficult dataset, i.e. weaker
signals. We align the stages of CL to the change in optimizer.
More details on the
training sets are given in Sec.~\ref{sec:training_sets}.
Alternative CL approaches avoid re-using the easier
dataset in further stages, but we decide to keep it also in the final stage so
that the model still remains robust to the stronger signals.

We train two CNN models with the same architecture, the same CL setup, etc.:
one on only-rectangular signals and one on only-exponential signals.
All training sets are equally split into noise and signal atoms.
The validation set,
used to compute the loss function after each epoch,
is $33\%$ of the training data.

\begin{table}[]
    \begin{tabular}{cccc}
    \hline
    \hline
                      &  & CL stage 1 & CL stage 2 \\ 
    \multirow{2}{*}{injection} & $\rho_{\textrm{inj}}$ & $U[6,40]$ & $\textrm{set}\,1+ U[4,10]$ \\ 
                    set  & $N_{\textrm{train}}$ & $4\times 10^4$ &
                      $\textrm{set}\,1+ 6\times 10^4$\\ 
                    optimizer  &  & Adam & SGD  \\ 
                    epochs  &  & 200 & 1000 \\ 
                    \hline
                    \hline
    \end{tabular}
    \caption{Parameters used for the two stages of curriculum learning (CL). We
    first train on strong signals with $\rho_{\textrm{inj}}$ randomly drawn from a uniform
    distribution $U[6,40]$, and then we add to the training weaker signals with
    $\rho_{\textrm{inj}}$ randomly drawn from $U[4,10]$. The number of samples we train on is
    $4\times 10^4$ in the first CL stage, and during the second CL
    stage we add $6\times 10^4$ more samples, this time with $\rho_{\textrm{inj}}$ in the weaker
    range. The final stage uses the full $10^5$ samples training dataset,
    including both high-SNR and low-SNR signals.
    The $\{t^0, \tau\}$ parameter ranges of the signals match the ones used for the
    search, which are listed in Tab.~\ref{tab:ephem}.
    }
    \label{tab:curriculum}
\end{table}

We start by training and testing on simulated Gaussian noise only,
but with realistic gaps matching those of the O2 run
(results will be shown in Sec.~\ref{sec:trainsynth_testsynth}). Real data, however, can present
instrumental artifacts which differ from a Gaussian background.
For tCW searches in particular, the disturbances that
most affect them are narrow spectral features similar to quasi-monochromatic CWs or tCWs
(so called ``lines'')
\cite{PhysRevD.97.082002, Keitel:2013wga, Keitel:2015ova}.
We also test these CNNs on real data
(Sec.~\ref{sec:trainsynth_testreal}). But as we will see, to be robust to non-Gaussianities of
the data one must use real data also in the training set. There are different
approaches that can be employed, e.g. only using real data in the training set,
or using a mixture of both simulated and real data. We explain our choice and
show our results in Sec.~\ref{sec:synth2real}.

\subsection{Training sets}
\label{sec:training_sets}
\subsubsection{Gaussian-noise synthetic data}
\label{sec:sets_synth}
For Gaussian noise training data, we use the approach of generating ``synthetic''
detection statistics (and atoms, see appendix A.4 of Ref.~\cite{Prix:2011qv}),
which avoids generating and analyzing SFT files
and therefore is considerably faster than full searches
over simulated Gaussian noise data.
Specifically, we use the
\texttt{synthesize\_TransientStats} program from \texttt{LALSuite} which draws
samples of the quantities needed to compute atoms,
either under the pure Gaussian noise assumption
or as expected for signals (with randomized parameters) in Gaussian noise.
It then produces the atoms
we need as input for the CNNs
and also the \Fstatmax and $\Btrans$ statistics
(for given transient parameter search ranges)
we will use for comparing the detection performance of the CNNs.

In each stage of the CL,
the training set is balanced, with half of the samples being pure noise and the
other half containing tCW signals. All signals correspond to Vela's sky location.
Frequency and spindowns do not matter for the synthesizing method.
The parameters $\cos\iota$, $\psi$, and $\phi_0$ are randomized over their natural ranges,
i.e. $\cos\iota\in[-1,1],\psi\in[0,\pi/2],\phi\in[0,2\pi]$.
The transient parameters of the signals
are drawn from the search ranges shown in Tab.~\ref{tab:curriculum},
which also summarizes the SNR ranges and set sizes for the two CL levels.

\subsubsection{Real data}
\label{sec:sets_real}
The real-data training set atoms are generated from analyzing a $0.1$\,Hz band
of LIGO O2 data \cite{LIGOScientific:2019lzm,O2SFTs}.
To avoid training bias,
we have to choose this band as disjoint from the search band around the nominal GW frequency of Vela,
but it should be close enough to have similar noise characteristics.
We center this band around $22.2$\,Hz, which yields a frequency region
without visible lines in the PSDs of the two detectors.

The noise samples are produced by running a grid-based search
with \texttt{PyFstat} \cite{Keitel:2021xeq, pyfstat}
and storing the output atoms,
using the same setup as in Tab.~\ref{tab:ephem}
except for the shift in frequency
and using a coarser frequency resolution
to reduce correlations between atoms at different $\Dop$:
namely \mbox{$df=5 \times 10^{-6}$\,Hz} for the first CL stage
and \mbox{$df=3 \times 10^{-6}$\,Hz} for the second.

The signal samples of this training set are generated by injecting
signals at random frequencies within the same offset band,
with spindowns fixed to the nominal GW values
(twice those from pulsar timing).
Atoms are then produced by a single-template \texttt{PyFstat} analysis
for each injection.

\subsection{Timings}
\label{sec:timings}
The training of the network takes about $1.8$\,hours on a Tesla V100-PCIE-32GB
for only synthetic data.
When including real data, the training takes twice as long in total. After
training, evaluation on the same GPU takes
\mbox{$c_{\textrm{CNN}} \approx 4\times10^{-4}$\,s} per sample, when averaged over a batch
of $10^4$ samples using Tensorflow's method \texttt{Model.predict\_on\_batch}.
This compares favorably with costs for the \Fstatmax or $\Btrans$ statistics, which
for the same set of transient parameters take $\approx
10^{-2}$\,s per sample for rectangular windows (similar with both the \texttt{LALSuite} CPU
code on a Intel Xeon Gold 6130 2.10GHz and the \texttt{PyFstat} GPU version from
Ref.~\cite{Keitel:2018pxz} on the same V100) and for exponential windows over 15\,s
on the CPU and \mbox{$c_{\Btrans^{\textrm{e}}}^{\textrm{GPU}} \approx 3\times
10^{-2}$\,s} on the V100.

\section{Evaluation method}
\label{sec:evaluation_method}
    The output $\rho_{\textrm{CNN}}$ can be used as a detection statistic for hypothesis testing.
We compare the performance of
the CNNs with other detection statistics using receiver operating characteristic
(ROC) curves on the separate test sets described below.
These show the probability of detection $p_{\textrm{det}}$, corresponding
to the fraction of signals above threshold, or true positives, as a function of the probability of false alarm
$p_{\textrm{FA}}$, which is equal to the fraction of false positives from pure-noise samples. The number of noise
samples determines how deep in $p_{\textrm{FA}}$ the curves can go, while the
number of signal samples determines the accuracy of the $p_{\textrm{det}}$
estimate. In the following we always use $\approx 10^7$ noise samples and $10^4$
signal samples.
The latter corresponds to $p_{\textrm{det}}$ uncertainties of $\lesssim 2\%$,
and we will consider differences in ROC curves below this level as marginally significant.
This disparity in noise and signal set sizes is due to the fact that, as mentioned
in Sec.~\ref{sec:search}, typical template bank sizes for standard searches can
reach the order of millions, and so the operating $p_\textrm{FA}$ at which we
assess the performance of our method must reach at least $10^{-6}$ to $10^{-7}$. 

We will compare two versions of the CNN,
one trained on only rectangular windows (output
$\rho_{\textrm{CNN}}^{\textrm{r}}$)
and the other trained on only exponential windows 
(output $\rho_{\textrm{CNN}}^{\textrm{e}}$),
with the detection statistics \Fstatmax and $\Btrans$.
We also use the r/e superscripts on these statistics
depending on the window assumed in their computation.
In addition, in this section we introduce
a two-stage filtering process combining a CNN with $\Btrans$.
\subsection{Testing sets}
Our test sets are constructed to match the O2 Vela search
discussed in Sec.~\ref{traditional_analysis}.
For both the synthetic and real data case, we generate two separate test sets:
both with the same $10^7$ noise samples,
corresponding to the number of $\lambda$ parameters in the template bank for our
reference search (compare Tab.~\ref{tab:ephem}).
But the $10^4$ signal samples are rectangular-shaped for
the first set, and exponentially-decaying-shaped for the second.
The generation methods for each set are the same as described in Sec.~\ref{sec:training_sets}.
The transient parameter ranges of the signal testing set are the same as those of the training sets,
while the $\rho_{\textrm{inj}}$ of the testing sets
are drawn from $U[4,40]$. The real data signal frequencies are randomly drawn from a $0.1$\,Hz
band around the nominal GW frequency of Vela.

\begin{figure}
    \scalebox{0.8}{
    \begin{tikzpicture}[node distance=2cm]
        \node (sfts) [process] {SFTs};
        \node (atoms) [process, below of=sfts,] {atoms};
        \node (Fmn) [process, below of=atoms, xshift=-2.5cm, yshift=-2cm] {$2\mathcal{F}(t^0,\tau)$};
        \node (cnn) [process, below of=atoms, xshift=2.5cm, yshift=-0.60cm] {$\rho_{\textrm{CNN}}$};
\node (logB) [process, below of=Fmn, yshift=-1cm]{$B_{\textrm{tS/G}}$};
        \node (detection) [decision, below of=cnn, yshift=-2.5cm,]{detection?};
        \node[inner sep=0,minimum size=0,left of=cnn, xshift=-3cm] (k) {}; 

        \draw [->, line width=0.4mm] (sfts) -- node[anchor=west]{matched filter}(atoms);
        \draw [->, line width=0.4mm] (atoms.west) -| node[anchor=east,
        yshift=-1.1cm]{partial sums}(Fmn.north);
        \draw [->, line width=0.4mm] (atoms.east) -| node[anchor=west, yshift=-1.1cm]{CNN}(cnn.north);
        \draw [->, line width=0.4mm] (Fmn) -- node[anchor=east]{marginalize}(logB);
        \draw [->, line width=0.4mm] (logB) -- (detection);
        \draw [->, line width=0.4mm] (cnn) -- (detection);
        \draw [-, line width=0.4mm, dashed] (cnn) -- node[anchor=south]{two-stage filtering}(k);
    \end{tikzpicture}
    }
    \caption{Flowchart of three different methods for the tCW search:
    from the SFT data to \Fstat atoms
    and then either via partial sums and marginalization to the $\Btrans$ statistic,
    or to the CNN.
    In the two-stage filtering method, after the CNN
    a subset of candidates get passed back to the branch leading to $\Btrans$.
    (For ``synthetic'' data, one starts directly from atoms.)}
    \label{fig:flowchart}
    \end{figure}
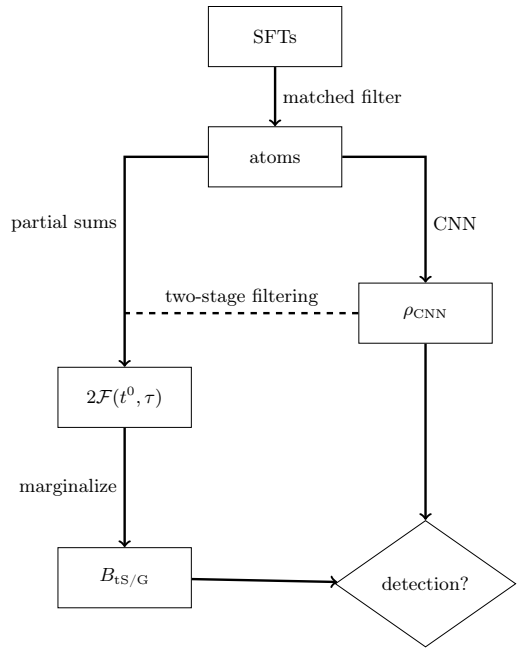

\subsection{Two-stage filtering}
\label{sec:pipeline}
We have seen in Sec.~\ref{sec:timings} that evaluating the CNNs is very fast.
In particular, for exponential windows it is faster than even the GPU implementation
of $2\mathcal{F}_{\max}^{\,\textrm{e}}$ and $\Btrans^{\textrm{e}}$
from Ref.~\cite{Keitel:2018pxz}.
Motivated by this, in all of the following tests we also evaluate a combined detection approach,
using the CNN as a preliminary filter stage
with $\Btrans$ as a second stage. The CNN can first be run
on all the available data, and then a follow-up with the traditional detection
statistic is done only if $\rho_{\textrm{CNN}}$ is above a given threshold. A
simplified flowchart of this method,
in comparison with the purely traditional and the pure CNN approaches,
is shown in Fig.~\ref{fig:flowchart}.

The amount of follow-up candidates can be set by choosing an operating
$p_{\textrm{FA}}^{\textrm{CNN}}$ for the CNN stage, so that we
gain as much sensitivity as possible while the whole pipeline is still
computationally less expensive than directly computing $\Btrans$ over all templates.

Towards low overall $p_{\textrm{FA}}$, we will see that
performance improves compared to the single-stage CNN
and approaches that of the pure $\Btrans$ more closely.
It is also possible for this two-stage method to achieve
higher $p_\textrm{det}$ at low $p_{\textrm{FA}}$ than pure $\Btrans$
if the CNN learns to better discard some noise features that would cause loud $\Btrans$ outliers.
On the other hand, the ROC
curves of this method will not arrive at the point $(p_{\textrm{det}} = 1,
p_{\textrm{FA}}=1)$ as is normally the case for single-stage detection methods.
Any signals lost, i.e. the amount of
$p_{\textrm{det}}$ lost, during the first stage cannot be restored, since only
the candidates passing that stage's threshold are passed on to $\Btrans$. Therefore, by
choosing a particular $p_{\textrm{FA}}^{\textrm{CNN}}$, the highest
probability of detection the two-stage method can reach will be at most
$p_{\textrm{det}}^{\textrm{CNN}}(p_{\textrm{FA}}^{\textrm{CNN}})$.

We choose
$p_{\textrm{FA}}^{\textrm{CNN}} = 10^{-3}$ as an operating point, since it offers a
good compromise between computational cost for the follow-up and the loss of
overall $p_{\textrm{det}}$ at that operating point, as we will see in
Sec.~\ref{sec:trainsynth_testsynth}.

In the case of an exponential window test set
with \mbox{$N \approx 1.15 \times 10^{7}$} templates
and with the hardware and timings as in Sec.~\ref{sec:timings},
the first stage takes \mbox{$c_{\textrm{CNN}} \times N \approx 1.3$\,hours}.
We pass a fraction of $10^{-3}$ of candidates to the second stage
where $\Btrans^{\textrm{e}}$ is evaluated.
This is then expected to take
\mbox{$c_{\Btrans^{\textrm{e}}}^{\textrm{GPU}}\times p_{\textrm{FA}}^{\textrm{CNN}} \times N \approx350$\,s.}
The two-stage filtering method thus takes in total less than
$1.5$\,hours against the \mbox{$c_{\Btrans^{\textrm{e}}}^{\textrm{GPU}}\times N \approx 95$\,hours} needed
without the first CNN stage. A single-stage $\Btrans^{\textrm{e}}$ search on a CPU (Intel Xeon
Gold 6130 2.10GH) would even take $\gtrsim 4\times10^4\,$hours.

\section{Performance on synthetic and real data}
\label{sec:results}
\subsection{Results on synthetic data}
\label{sec:trainsynth_testsynth}
\begin{figure}[t]
    \includegraphics[width=\columnwidth]{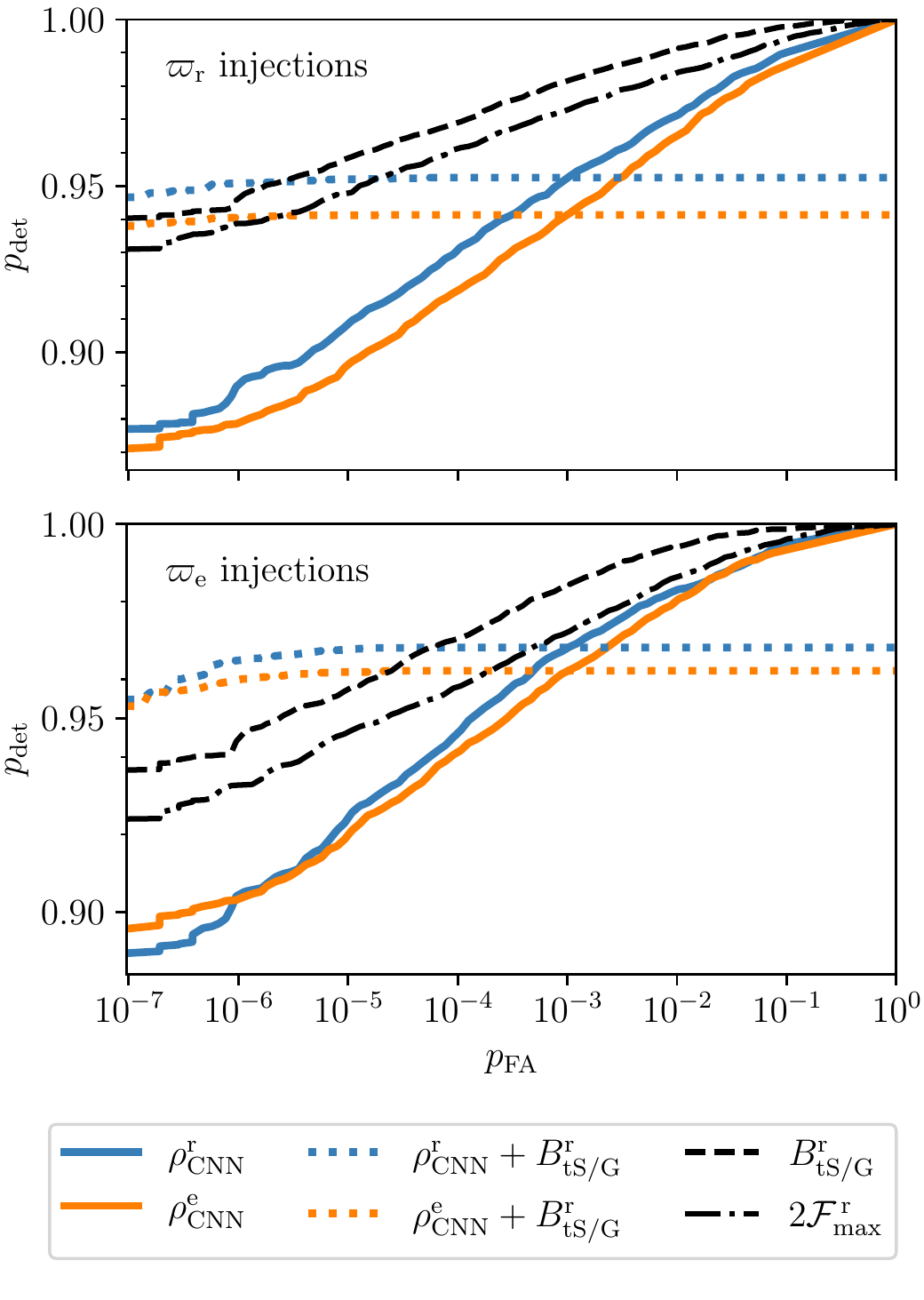}
    \caption{ROC curves with synthetic data
    as testing set and CNNs trained on synthetic
    data. Note the vertical axis is zoomed in. Both panels use the same noise
    testing set, but the signals in the testing sets are rectangular-windowed in the
    upper panel, and exponentially decaying in the bottom panel. For both
    panels, the black lines are the ROC curves for $2\mathcal{F}_{\max}^{\,\textrm{r}}$ (dash-dotted) and
    for $\Btrans^{\textrm{r}}$ (dashed), always searching for a rectangular signal (even when
    the test set contains exponentially decaying signals). The solid windowed
    lines are the ROC curves for the CNNs, trained on only rectangular-shaped
    signals (blue) and on onlys exponentially decaying signals (orange). The
    dotted colored lines are the ROC curves for the two-stage filtering method,
    with the corresponding windows used in training.}
    \label{fig:roc_synth}
\end{figure}

We start by testing the CNNs that were trained on synthetic data only,
also using purely synthetic data for the test. As mentioned in
Sec.~\ref{sec:sets_synth}, synthetic data are random draws of the \Fstat
atoms, assuming an underlying Gaussian noise distribution, from which the
tCW detection statistics $2\mathcal{F}_{\max}^{\,\textrm{r}}$ and $\Btrans^{\textrm{r}}$
can then be calculated.\footnote{
We omit the much more expensive $2\mathcal{F}_{\max}^{\,\textrm{e}}$ and $\Btrans^{\textrm{e}}$ statistics in this section.
We will only compute them on a full test set for comparison once, on real data, in Sec.~\ref{sec:synth2real}.}
The detection problem is easier for this case than for real data,
so we show these results as a starting point.
In Fig.~\ref{fig:roc_synth} we show ROC curves comparing the CNNs
against the other two detection statistics.
The two subplots refer to the two different test sets, which differ only in the
window function of the injected signals, but for both cases the standard
detection statistics use rectangular windows.

As expected \cite{Prix:2011qv}, $\Btrans^{\textrm{r}}$ performs marginally better than
$2\mathcal{F}_{\max}^{\,\textrm{r}}$. Both their probabilities of detection surpass
\mbox{$p_{\textrm{det}}=90\%$} even at $p_{\textrm{FA}}$ as low as $10^{-7}$.
The two CNNs (trained on only rectangular and only exponentially decaying windows)
perform similarly overall,
with at most $\lesssim 1\%$ difference in $p_{\textrm{det}}$
between each other
and a loss with respect to $\Btrans^{\textrm{r}}$ of
at most $7\%$ for rectangular
and $5\%$ for exponential windows.

The two-stage filtering reaches a plateau at higher $p_{\textrm{FA}}$ values,
corresponding (as mentioned in Sec.~\ref{sec:pipeline}) to the fixed
\mbox{($p_{\textrm{FA}}=10^{-3},p_{\textrm{det}}<1)$} operating point of the
first stage of the method.
However, at low $p_{\textrm{FA}}$ it considerably improves $p_{\textrm{det}}$ for both CNNs.
It actually performs marginally better than $\Btrans^{\textrm{r}}$
below \mbox{$p_{\textrm{FA}}<10^{-5}$} for both types of signals:
by $\approx1$\% for rectangular and $\approx2$\% for exponentially decaying injections.
\begin{figure*}[t]
    \includegraphics[width=\textwidth]{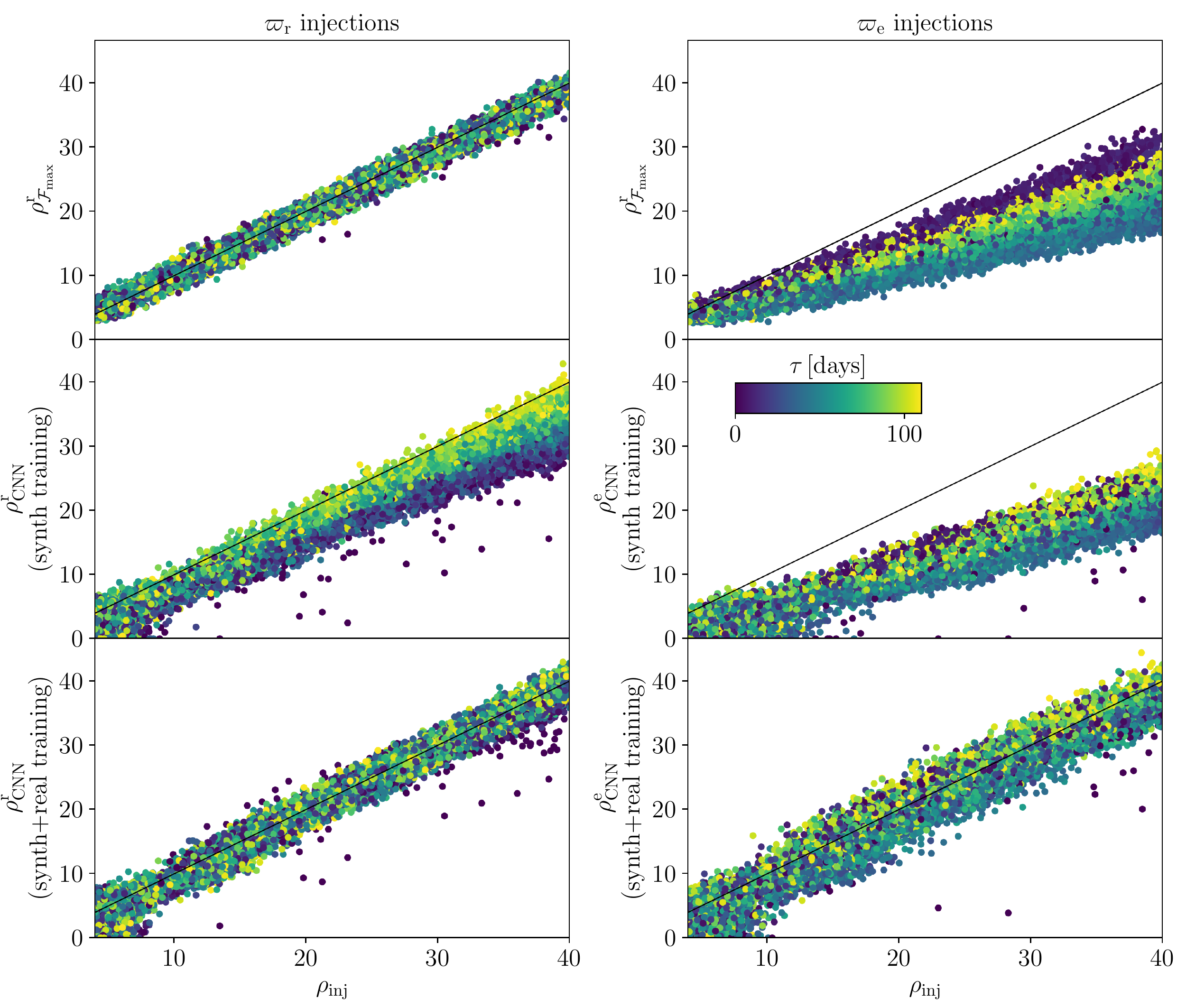}
    \caption{Estimated SNRs from three different methods
    (one for each row) as a function of injected SNRs in real LIGO O2 data.
    From top to bottom, the
    estimated SNRs are $\rho_{\mathcal{F}_{\max}}^{\,\textrm{r}}$ estimated from
    $2\mathcal{F}_{\max}^{\,\textrm{r}}$, in the second row
    $\rhocnn^{\textrm{r/e}}$ is estimated from the CNNs trained on
    only synthetic data with windows matching that of the test set and lastly,
    in the third row $\rhocnn^{\textrm{r/e}}$ is estimated from the
    CNNs trained on both synthetic and real data. The injected signals are
    rectangular-shaped on the left column and exponentially decaying on the
    right column. The color scale (shared between all panels)
    indicates the durations of the injected signals.
    For guidance, we always plot the diagonal of the
    subplots, where ideally the points should align.}
    \label{fig:SNR_panel}
\end{figure*}

The improvement is not necessarily significant compared with counting uncertainty,
and in the exponential case where it seems larger we have also not compared
to the traditional statistics using exponential windows.
Still, in general such an improvement is indeed possible
because $\Btrans$ is not an optimal statistic in either case.
As discussed in Ref.~\cite{Prix:2009tq} (for CWs), a better detection statistic for a given signal population
can in general be obtained by marginalization over an appropriate amplitude prior,
as opposed to the unphysical prior that the maximization used for the \Fstat
(and derived statistics such as \Fstatmax and $\Btrans$)
corresponds to.
In addition, Ref.~\cite{PhysRevD.105.124007} has demonstrated how in particular for short data stretches
a more sensitive statistic than the standard \Fstat can be constructed,
mainly exploiting a reduction in noise degrees of freedom.
Therefore, it seems consistent that a CNN can also learn to behave better than the standard
statistics, and more similarly to the improved alternatives, in at least certain
parts of the tCW parameter space.

\subsection{Results of synthetic-trained CNNs on real data}
\label{sec:trainsynth_testreal}
In the previous section we have shown how CNNs can be used as a competitive
method against the standard detection statistics, at least in the context of
synthetic data.
We now evaluate the same models trained with synthetic data,
but now we test on real data. 
The corresponding CNN detection probabilities suffer additional losses,
compared to Fig.~\ref{fig:roc_synth},
of up to $7\%$ and $16\%$ for rectangular and exponential windows respectively.
Instead of showing these ROC curves,
we proceed directly to identifying and addressing the main reason for these losses.

For real-data injections,
we show in Fig.~\ref{fig:SNR_panel} the estimated $\rhocnn$
compared with the traditional estimator~\cite{Prix:2015cfs}
\begin{equation}
    \rho_{\mathcal{F}_{\max}}^{\textrm{r}} = \sqrt{2\mathcal{F}_{\textrm{max}}^{\,\textrm{r}} -4} \,,
    \label{eq:rho_2Fmax}
\end{equation}
both as functions of $\rho_{\textrm{inj}}$.
Different injection windows are covered in the two columns. Ideally, each estimator would
align all the points on the diagonals of these plots.

We see that $\rho_{\mathcal{F}_{\max}}^{\textrm{r}}$ (first row) aligns well with the
injected $\rho_{\textrm{inj}}$ for rectangular windows, but for exponentially decaying
signals there is a loss in recovered SNR (i.e. the points do not align with the
diagonal) because the injected window (exponential) and the search window
(rectangular) do not match. The loss in SNR is also enhanced by a 12-day gap in
the data, which we discuss in more detail in Appendix~\ref{sec:gaps}. 

In the second row, we see that the CNN trained on synthetic data
and rectangular windows recovers $\rho_{\textrm{CNN}}^{\textrm{r}}$
that mostly align with $\rho_{\textrm{inj}}$, but the loss
increases as the duration of the injected signal
decreases. The loss in SNR is
even more evident for $\rho_{\textrm{CNN}}^{\textrm{e}}$ on exponentially decaying signals.
Since both of these CNNs were
trained on synthetic data only, it is natural to expect that they are not robust
to detector artifacts that could contaminate the data, especially at shorter durations
\cite{LIGO:2021ppb,Keitel:2015ova}. We will now demonstrate that it helps to include real data
in the training set to mitigate this effect.

\subsection{Performance of training on real data}
\label{sec:synth2real}
\begin{figure}
    \includegraphics[width=\columnwidth]{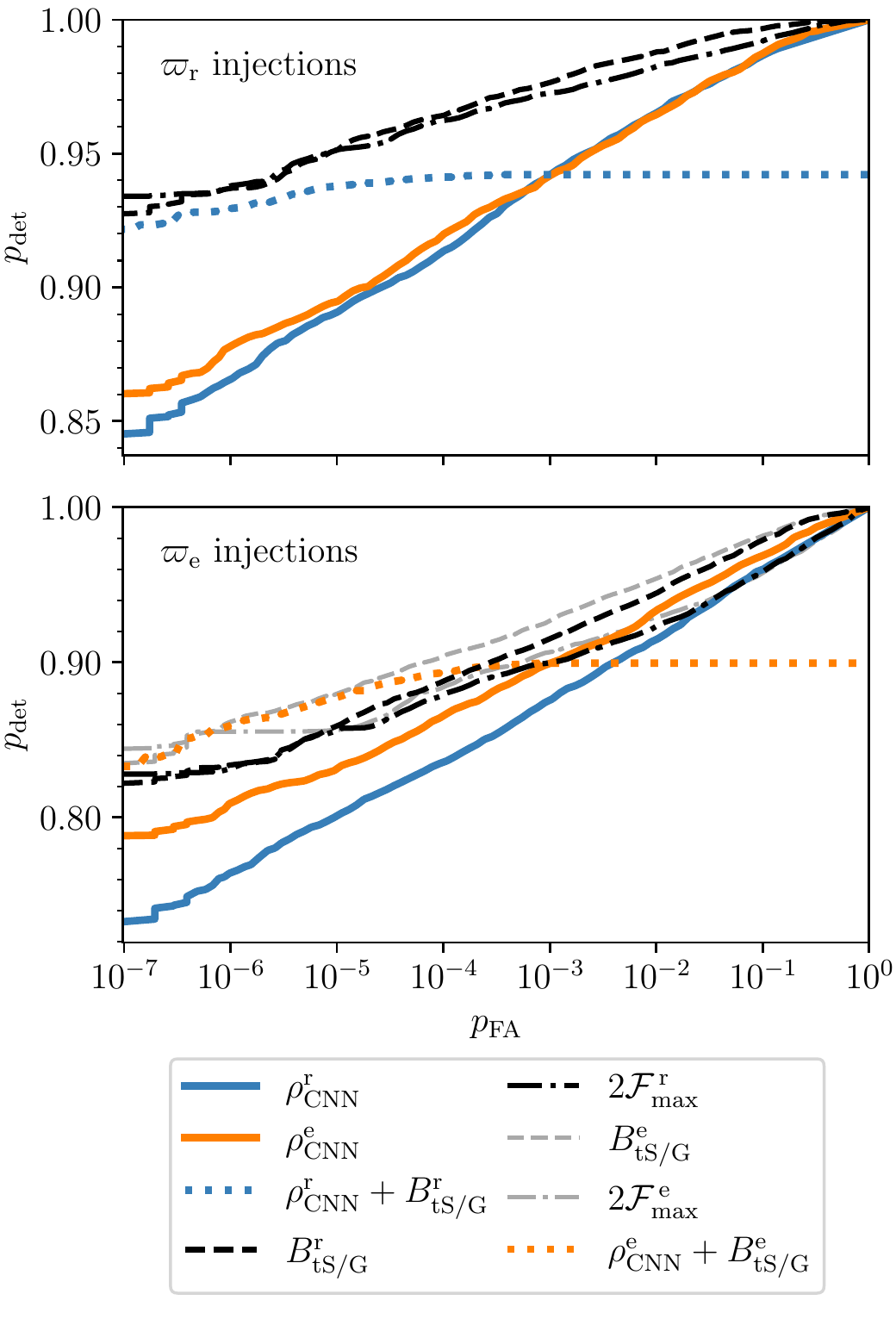}
    \caption{ROC curves with real data as the
    testing set and CNNs trained on both synthetic and
    real data. Note the vertical axis is zoomed in. Both panels use the same
    noise testing set, but the test signals are rectangular-shaped in
    the upper panel, and exponentially decaying in the bottom panel. The legend
    structure is the same as in Fig.~\ref{fig:roc_synth}, with the
    addition of two extra ROC curves in the exponential test case corresponding
    to the $2\mathcal{F}_{\max}^{\,\textrm{e}}$ and $\Btrans^{\textrm{e}}$
    statistics computed with the more expensive exponential window.}
    \label{fig:roc_synth2real}
\end{figure}

To improve sensitivity, we also trained CNNs using
real data, keeping the same architecture.
As mentioned before in Sec.~\ref{sec:training_strategy}, there are different ways one
can incorporate real data in the training set. We have tried different
implementations using the same total number of training epochs: training a model from scratch
on only real data, training a model on a mixture of synthetic and real data, and
taking the previous synthetic-only-trained model and continuing
its training on only real data (similar to a CL approach).

We find that training
exclusively on real data yields a good signal recovery, i.e. removes the SNR losses seen
in Fig.~\ref{fig:SNR_panel}, but overestimates the SNR of the noise samples,
leading to inferior ROC results overall. This effect is mitigated when training
with both synthetic and real data. Between
the two options of training on a mixture of both all at once, and
first training on synthetic and then on real data, the latter
performs best. More precisely,
when first training on synthetic and then on real data, we take the previously
trained CNNs on synthetic data for 200+1000 epochs (first CL stage + second CL
stage) and continue their training on only real data for another 200+1000 epochs.

As before, we train two models,
each on a single window type. In the third row
of Fig.~\ref{fig:SNR_panel} we show the recovered $\rhocnn^{\textrm{r/e}}$
from these networks (trained first on synthetic and then real data)
on real-data injections.
For rectangular injections, the loss in SNR is now
reduced compared to the CNNs with only synthetic training data.
The issue of more SNR loss for shorter signals is
largely resolved except for a few underestimated outliers with short durations:
the CNN is now able to recover both long and short signals well. In the exponential
case, $\rhocnn^{\textrm{e}}$ also aligns well with the injected $\rho_{\textrm{inj}}$,
with just a wider scatter.

The models trained on both synthetic and real data also perform well on real noise
samples, not producing overly loud outliers. ROC curves for this case are shown
in Fig.~\ref{fig:roc_synth2real}. For rectangular signals, sensitivities of
all the different statistics are quite similar to those on synthetic data
(top panel of Fig.~\ref{fig:roc_synth}). However, while for synthetic data the
two-stage filtering performance matched or exceeded that of the standard
detection statistics at low $p_{\textrm{FA}}$,
it now shows a marginal loss in the same regime with respect to the standard
statistics, but of only $\approx 1\%$.

Also, the CNN trained on exponential windows performs marginally better even with
rectangular window injections at the lowest values of $p_{\textrm{FA}}$. It was
noted by Ref.~\cite{Prix:2011qv} that also
$2\mathcal{F}_{\max}^{\textrm{\,e}}$ can outperform $2\mathcal{F}_{\max}^{\textrm{\,r}}$ on
rectangular signals. However, without larger studies on different configurations
and data sets it cannot be determined if the observed phenomenon in the CNNs is
related or just due to specifics of the training and test data sets.

On the other hand, for the exponential test set, $p_{\textrm{det}}$ for all
statistics has systematically worsened by $10\%$ to $15\%$ compared to the
synthetic results from Fig.~\ref{fig:roc_synth}. This could be because the weak
late-time portions of the signals are more difficult to pick up in real noise.
The CNNs lose relatively less detection power, meaning that the gap in
$p_{\textrm{det}}$ against the traditional statistics has narrowed down: at the lowest
$p_{\textrm{FA}}$, the difference in $p_{\textrm{det}}$ between
$\Btrans^{\textrm{r}}$ and $\rhocnn^{\textrm{e}}$ is down to $3\%$. 

Since this is our most complete test set, we also show the results of the more computationally expensive statistics computed with windows matching the injections, i.e. $2\mathcal{F}_{\max}^{\,\textrm{e}}$ and
$\Btrans^{\textrm{e}}$. To obtain these we used the GPU implementation
from Ref.~\cite{Keitel:2018pxz}. Their ROC curves improve by $2-3\%$ over the statistics with
rectangular windows. 

For \mbox{$p_{\textrm{FA}}<10^{-5}$}
the two-stage filtering ROC curve very closely matches that of $\Btrans^{\textrm{e}}$.
It cannot yield higher $p_{\textrm{det}}$ (as it did in the synthetic case)
on this specific data set,
because the template corresponding to the loudest $\Btrans^{\textrm{e}}$ outlier
also passes the first-stage $\rho_{\textrm{CNN}}^{\textrm{r/e}}$ threshold.
But it does converge to the sensitivity of the traditional statistic
when far enough to the left of the ROC to go below the plateau level
set by the first stage.
As discussed before, it also takes 80 times less time even with the GPU implementation of Ref.~\cite{Keitel:2018pxz}.

It is important to remember that ROC performance depends on the population of
signal candidates considered. In this case we use signals with
$\rho_{\textrm{inj}} \in [4,40]$. In Sec.~\ref{sec:upper_limits} we will revisit
the same real-data noise set but with a different injection set matching the one
used in Ref.~\cite{Keitel:2019zhb} for establishing upper limits.

\section{Parameter estimation with CNNs}
\label{sec:parameter_estimation}
\begin{figure}
    \includegraphics[width=1\columnwidth]{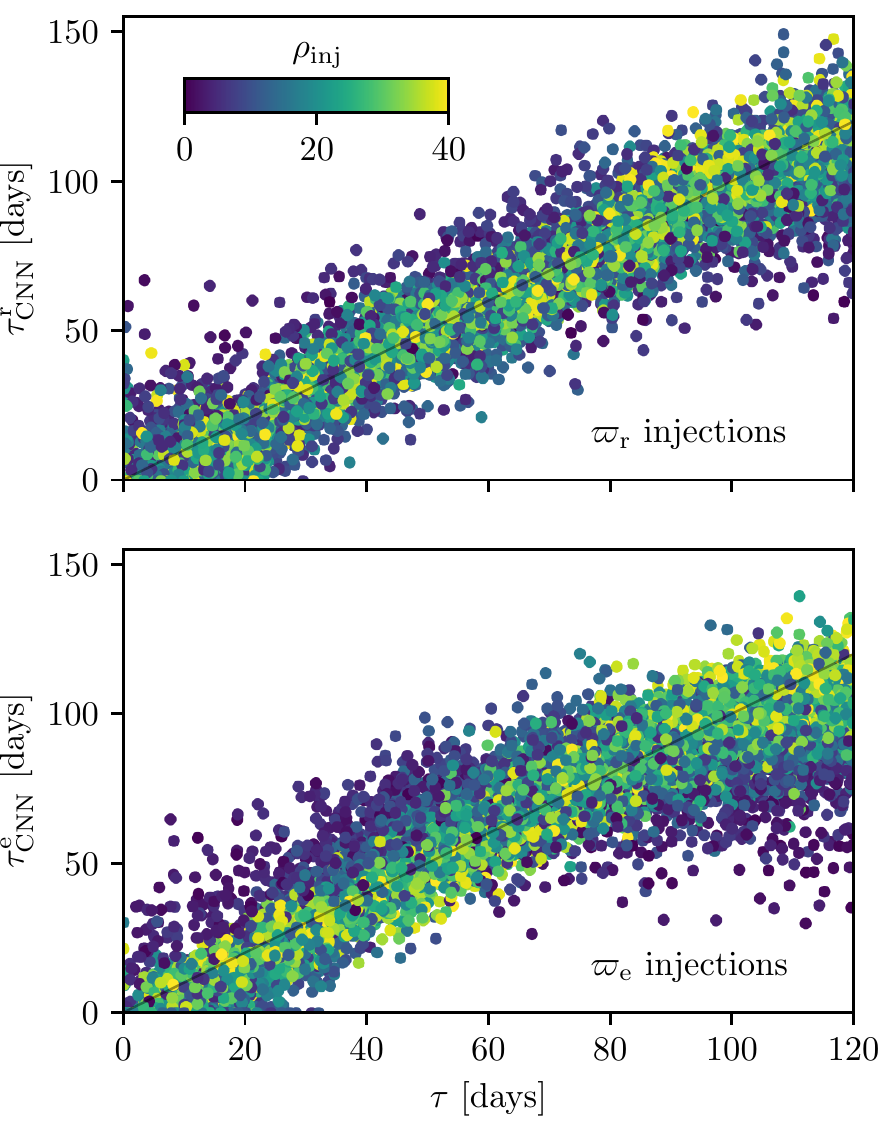}
    \caption{Duration parameters $\tau_{\textrm{CNN}}^{\textrm{r/e}}$ estimated by the two-output CNNs
    trained on rectangular (top panel) and exponential signals (bottom panel),
    plotted against the duration $\tau$ of the injected signal. 
    These use the same synthetic test sets as
    in Sec.~\ref{sec:trainsynth_testsynth},
    with windows matching the training in each panel.
    Only signals passing the
    $\rho_{\textrm{CNN}}^{\textrm{r/e}}$ thresholds
    of the two-stage filtering are plotted.
    The color scale (shared between both panels) corresponds to the
    $\rho_{\textrm{inj}}$ of the injected signals.}
    \label{fig:PE}
\end{figure}

So far, we have considered a CNN with only one output: the estimated SNR,
which only allows yes/no detection statements.
In practice, one will also be interested in the parameters of a signal candidate.
Estimates of the frequency-evolution parameters $\lambda$
are naturally obtained from where the detection statistic
peaks in the template bank.
In our preferred setup, the two-stage filtering, final candidates
will additionally have $(t^0,\tau)$ estimates from the algorithm in Ref.~\cite{Prix:2011qv}.

However, CNNs can also estimate multiple parameters directly.
In this section, we present a first exploration of this possibility via training CNNs with multiple outputs.
We keep the same architecture as before
with only the addition of one extra node to the output
layer: we now want to output both $\rho_{\textrm{CNN}}$ and a duration estimate $\tau_{\textrm{CNN}}$ for a potential signal.
One could also
estimate the starting time $t^0$ by adding another output, but since in our setup the
starting time is limited to a small range around  $T_{\textrm{gl}}$, we
ignore it for this first proof of concept.

We here show the results of two newly trained models with the additional $\tau_{\textrm{CNN}}$
output and trained and tested with synthetic data. The duration input labels for
training are normalized to a range between $[0,1]$. 
The output can then be transformed back to the
duration in days. In general, we find that the accuracy of the $\rho_{\textrm{CNN}}$ estimation
remains unaltered, while the quality of $\tau_{\textrm{CNN}}$ estimates depends on which type of windows
we use in training. These are shown for one CNN trained on only rectangular
signals and by one trained on only exponential signals, as functions of
the injected durations from the test set, in
Fig.~\ref{fig:PE}.
To focus on signals that are at least marginally detectable,
we only include those passing the  $\rho_{\textrm{CNN}}^{\textrm{r/e}}$ thresholds
of the two-stage filtering,
in both cases corresponding to \mbox{$p_{\textrm{FA}}^{\textrm{CNN}} = 10^{-3}$}.

For the rectangular-trained CNN, the estimated durations $\tau^{\textrm{r}}_{\textrm{CNN}}$ follow quite well the true durations, 
but with some scatter especially for low-SNR signals.
About 3\% of signals form a set of outliers with durations underestimated
by over 95\%, mostly for shorter injections.
Excluding these outliers, the remaining distribution of relative errors
is well-centered around zero
with a root-mean-square (RMS) of $27\%$. For the exponential-trained CNN,
there is a mild trend towards more under-estimated $\tau^{\textrm{e}}_{\textrm{CNN}}$
values for longer injected signals.
In this case, about 7\% of signals are underestimated by over 95\%,
but after excluding this subset the error distribution is only slightly wider
with a RMS of $29\%$.

We also tested these synthetic-trained CNNs on real data. For rectangular
signals, the duration estimation is overall quite robust, and the plot
equivalent to Fig.~\ref{fig:PE} follows the same shape except for some more cases of
overestimation at high durations. The outlier percentages for real data
estimations slightly increase to 6\% and the RMS error stays close to 27\%. For
exponential signals, there is noticeably more overestimation and the outlier
percentage increases to 11\% and the RMS error to 47\%. 

As an additional investigation, we trained a two-output CNN on signals with \textit{both}
rectangular and exponential windows. Overall it behaves like a mixture of the
results of the two separately trained networks, with the exception
of stronger overestimation at long durations and high SNRs in real data.
Performance could potentially be improved by adding another output acting as a
window label, since the duration parameter $\tau$ actually has different
meanings for the two different windows, as also discussed in
Ref.~\cite{10.1093/mnras/stac3665}. This would be similar to how Ref.~\cite{Prix:2011qv}
has already discussed treating window function choice as an additional parameter
in Bayesian parameter estimation.

This and other possible improvements to the architecture and training of the
two-output network to obtain better accuracy, and any further extensions of this
approach to parameter estimation with CNNs, are left for future work.

\section{O2 Vela glitch upper limits}
\label{sec:upper_limits}
\begin{figure*}[th]
    \includegraphics[width=1\textwidth]{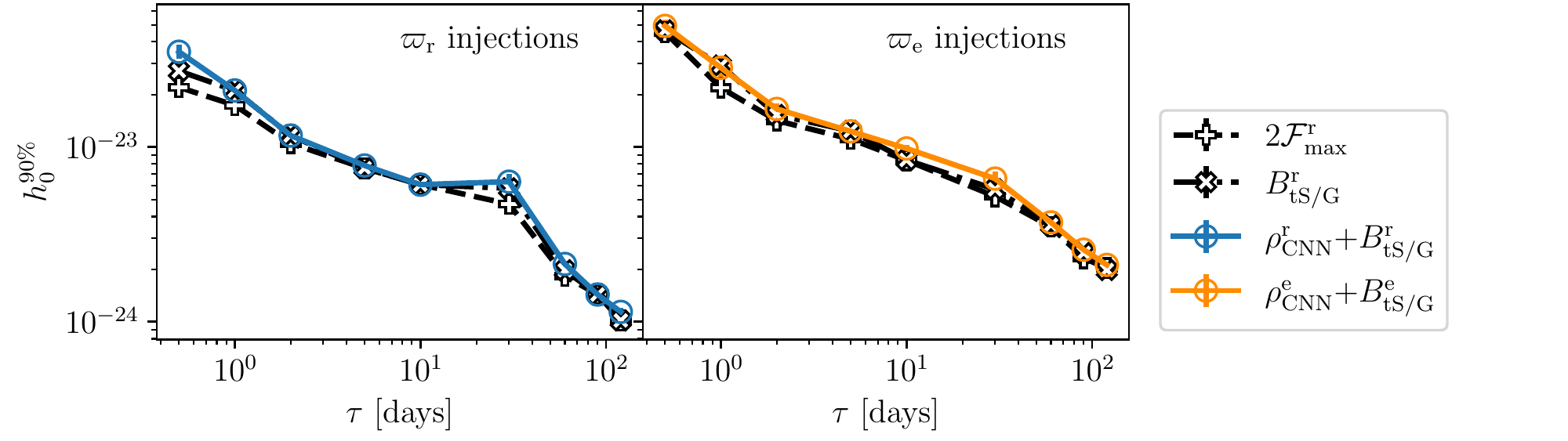}
    \caption{Upper limits on $h_0$ as a function of duration $\tau$ for the different methods: $2\mathcal{F}_{\max}^{\,\textrm{r}}$ (dash-dotted black) and $\Btrans^{\textrm{r}}$ (dashed black), both recovering rectangular signals. The two different shapes of the injections are used in the different panels (rectangular on the left, exponential on the right). In each subplot we also show the two-stage filtering trained on only rectangular-shaped signals (left panel, solid blue) and two-stage filtering trained on only exponentially decaying signals (right panel, solid orange). }
    \label{fig:upper_limits}
\end{figure*}
No detection of a post-glitch tCW signal was reported in the O2 search
\cite{Keitel:2019zhb}, so upper limits on the GW strain amplitude were set.
Having found no interesting outliers in the CNN results on the same data set either,
we repeat the procedure here, but covering both rectangular and exponential window choices.

Upper limits are computed by injecting simulated signals in the same data used in the
search and then counting how many signals are recovered by the chosen method.
For a set of injections at fixed duration $\tau$,
one can then fit a sigmoid curve of the counts against
injected amplitude $h_0$ to find the upper limit amplitude $h_0^{90\%}$ at which
$90\%$ of the injected signals are recovered above the threshold of each
statistic. 
A more detailed explanation can be found in the appendix of Ref.~\cite{Keitel:2019zhb}.

As the thresholds to distinguish between noise and signal
candidates we use the highest output of the search
over the original data and the ranges given in Table~\ref{tab:ephem},
for each detection statistic. This
means that the thresholds of the different detection statistics
do not necessarily correspond to the same $p_{\textrm{FA}}$, but it is a consistent method that
can be applied to any statistic.\footnote{Alternatively, one could use the
method described in Ref.~\cite{Tenorio:2021wad} to estimate the distribution of the expected loudest outlier from the background and
derive a threshold. However, we find $\rhocnn$ to show more complicated
distributions on our test data than the typical ones for \Fstatmax and $\Btrans$. So,
further investigation would be required to apply the method.}
It is also a conservative choice in the sense that any lower threshold would
produce stricter upper limits, but would require outlier followup. In the
special case of the two-stage filtering, the first threshold on the CNN stage is
chosen to let a fraction $10^{-3}$ of candidates pass and the final threshold is
given by the highest $\Btrans$ in that remaining set.

We create an injection set with simulated signals of amplitude $h_0$ in the
ranges used in the O2 search \cite{Keitel:2019zhb}, chosen to
correspond to detection statistics around the threshold values. This translates to
mostly weaker signals compared to the test sets
described in Sec.~\ref{sec:evaluation_method}
and used for the ROC curves in Sec.~\ref{sec:results}.
The durations of the injections are not distributed uniformly as done
for the previous testing sets, but rather chosen at discrete steps from
0.5 to 120\,days.

In Fig.~\ref{fig:upper_limits}, we reproduce the plot from Ref.~\cite{Keitel:2019zhb}
for rectangular-window signals,
showing the upper limits on $h_0^{90\%}$ as a function of the duration parameter $\tau$,
and we extend it with an exponential-window injection set.
We show upper limits derived from several methods discussed before, namely
$2\mathcal{F}_{\max}^{\,\textrm{r}}$, $\Btrans^{\textrm{r}}$,
and the two-stage filtering method using CNNs trained on only
rectangular windows or only exponential windows, combined with $\Btrans^{\textrm{r/e}}$.
The statistics $2\mathcal{F}_{\max}^{\,\textrm{e}}$ and $\Btrans^{\textrm{e}}$
for the full injection set 
are again omitted due to computational cost.

The upper limits from
$2\mathcal{F}_{\max}^{\,\textrm{r}}$ using rectangular injections can be directly compared to those
obtained in Ref.~\cite{Keitel:2019zhb}. They are consistent within $1\sigma$ error bars, and the small differences are due to the
different individual injections and the different sigmoid fitting procedure and
uncertainty estimation
(matching the implementation as in Ref.~\cite{LIGOScientific:2021quq}).

The upper limits from the two-stage filtering reach values close to $2\mathcal{F}_{\max}^{\,\textrm{r}}$ and $\Btrans^{\textrm{r}}$,
higher only by 6--7\% for both signal types.

As seen before in the real-data ROC curves,
the two-stage filter cannot improve over $\Btrans^{\textrm{r/e}}$
on these specific data sets
because the templates corresponding to the loudest outliers from those statistics
also pass the first-stage $\rho_{\textrm{CNN}}^{\textrm{r/e}}$
thresholds.

Another factor that could potentially be relevant in this new, typically weaker
injection set, is that upper limits depend strongly on the performance for weak
signals near threshold, which are the most challenging for CNNs.
This is exacerbated by using steps in the amplitude $h_0$, instead of SNR, to
quantify the strength of the signals. While the SNR takes into account the other
parameters that affect detectability, especially the inclination $\cos\iota$,
the amplitude does not contain this information
and so cannot be used as a direct proxy for the SNR.
Therefore, at each $h_0$ step of the injection set,
there can be a tail of signals with low SNRs,
where the differences between the CNN and traditional detection statistics are more significant.

\section{Conclusions}
\label{sec:conclusions}
In this work we present a new method based on CNNs for detecting potential
tCWs -- long-duration quasi-monochromatic GW signals --
from glitching pulsars. CNNs have proven to be promising tools for detecting
various GW signals, but have not been tested before on tCWs from glitching pulsars.
Previous searches were entirely matched-filtering based and limited in
computational cost. In this work we have used intermediate matched-filter
outputs (the \Fstat atoms) as input to the CNNs, which allows to replace the most
computationally expensive part of the analysis. In particular, practical
searches with the \Fstatmax and $\Btrans$ statistics were limited to
assuming constant amplitude (rectangular window) tCWs due to the much higher
cost of other window functions, while with CNNs we can easily train on different
windows. 

The CNNs are constructed to output an estimator of SNR in the data. We have trained
and tested CNNs for either rectangular or exponentially decaying windows, first
on synthetic, Gaussian data, but with gaps corresponding to the timestamps of
the LIGO O2 data after the Vela glitch of 2016. We use curriculum learning, i.e.
first train on stronger and then also on weaker signals. Then we have tested
these CNNs on the real LIGO data from O2 as previously analyzed in Ref.~\cite{Keitel:2019zhb}, and also trained with real data using the same
architecture to improve the results. We find the best results when starting
from the model we had trained on synthetic data and continuing its training on
only real data. 

We find that a simple implementation of such atoms-based CNNs already approaches
within 10\% of the detection probability at fixed false-alarm probability
of the traditional detection statistics. As a CNN-based method that
mostly closes this remaining gap, we propose a two-stage filtering method
consisting of first applying the CNN to all signal templates and only passing
candidates above a certain threshold on $\rho_{\textrm{CNN}}$ to the $\Btrans$
statistic. For a real data test set containing injected signals of broad SNR
range (from 4 to 40), the probabilities of detection at false-alarm probabilities as low as
$10^{-7}$ are only $\lesssim 2\%$ lower than those of $\Btrans$
for rectangular windows and 4\% better for exponential windows.
Comparing the computing time of the two-stage filtering with that of $\Btrans$
for exponential signals using the GPU implementation of Ref.~\cite{Keitel:2018pxz},
we find that our method is $80$ times faster than evaluating the full data
without the CNN stage.

We then use this new method to 
set updated observational upper limits on the GW strain amplitude $h_0$ after the O2 Vela glitch,
with extended parameter space coverage
including exponentially decaying signals.
For this we make a separate injection set with signal parameters in the same ranges as in Ref.~\cite{Keitel:2019zhb}.
The upper limits from the
two-stage filtering almost match those reached by the standard statistics, 
higher only by $\lesssim7\%$.

Thanks to its computational efficiency, the CNN-based approach
can be a competitive method for the overall tCW search effort
that is also complementary to the reference method:
if extended to more generic searches,
it will help increase overall detection probability by extending discovery space
while one can afterwards still brute-force evaluate the traditional detection statistics
when interested in pushing for the deepest possible upper limits on individual targets.

Considering such further applications of CNNs to tCWs from glitching pulsars,
we have here focused on a single target pulsar and data set,
but the approach can be generalized to other targets or even unknown sky positions,
either by training from scratch in each case or through transfer learning \cite{Gupta_2022}.
Due to the computational efficiency of both the training and
evaluation phase, the two-stage filtering could thus be scaled up to broader searches
covering a larger variety of targets than currently feasible with the standard methods,
especially when wanting to include multiple amplitude evolution window options.

Also, since CWs can be obtained from the tCWs model by setting a rectangular
window function to cover the entirety of the data, this method can easily be
applied to persistent CW targeted searches. Our implementation of CNNs, however,
was designed specifically to avoid the computationally expensive computation of
partial sums of the \Fstat atoms over all the combinations of the transient parameters, which does
not concern CW searches.

Furthermore, one could go beyond the current configuration, in which we trained the CNNs on \Fstat atoms, i.e. quantities
computed during the matched filtering step. This still
constrains the frequency evolution of the signal to be CW-like, but already
allows for flexible amplitude evolution and significant speed-up compared to the
traditional method, effectively allowing to search a wider parameter space at
the same cost. A different approach would be to train a CNN
(or different type of network)
on spectrograms or the full timeseries strain data, which could allow searches for unmodeled tCWs
both in frequency and amplitude evolution.

The major drawback would be that the
amplitudes of tCW signals are expected to be very weak, with $h_0$ upper limits from
O3 reaching $10^{-25}$ \cite{LIGOScientific:2021quq}. Such signals are too weak to be directly discernible, e.g. in time-frequency maps of the data. A study of using Fourier
transforms of the data as input to a CNN to search for persistent CWs was
done in Ref.~\cite{2019PhRvD.100d4009D}, and a broader set of machine-learning strategies
on SFTs was evaluated in a public Kaggle challenge \cite{G2NetCWKaggleChallenge}.
Similar approaches could be applied to tCWs as well,
but it will be a challenging problem, which we leave for future work.
Despite not reaching the high
sensitivities of purely matched filtering methods, these faster, generic
searches could still extend the tCW science case
by enabling the search for post-glitch GWs corresponding to more general glitch
models and over larger parameter spaces, potentially leading to all-sky, all-time,
all-frequency blind searches of tCWs.

\section*{Acknowledgements}
We thank Vincent Boudart for making the CNN architecture plot,
the UL\`iege group from the STAR Institute, in particular Jean-Ren\'e Cudell,
Maxime Fays, Gr\'egory Baltus and Prasanta Char 
for hosting L.M.M. during a g2net (CA17137)
short term scientific mission working on this project,
Karl Wette for
valuable help with LALSuite SWIG wrappings~\cite{Wette:2020air},
Xingyu Zhong and Andrew L. Miller for initial exchanges about applying CNNs to tCWs,
Pep Covas for an interesting seminar talk on his improved short-segment detection statistic
that gave us useful inspiration on interpreting our ROC results,
and Alicia M. Sintes, Maite Mateu-Lucena and other members of the LIGO--Virgo--KAGRA
Continuous Waves working group for useful comments. This work was supported by
the Universitat de les Illes Balears (UIB); the Spanish Ministry of Science and
Innovation (MCIN) and the Spanish Agencia Estatal de Investigacio{\'o}n (AEI)
grants PID2019-106416GB-I00/MCIN/AEI/10.13039/501100011033; the MCIN with
funding from the European Union NextGenerationEU (PRTR-C17.I1); the FEDER
Operational Program 2021-2027 of the Balearic Islands; the Comunitat
Aut{\`o}noma de les Illes Balears through the Direcci{\'o} General de
Poli{\'i}tica Universitaria i Recerca with funds from the Tourist Stay Tax Law
ITS 2017-006 (PRD2018/24, PRD2020/11); the Conselleria de Fons Europeus,
Universitat i Cultura del Govern de les Illes Balears; and EU COST Actions
CA18108 and CA17137. DK is supported by the Spanish Ministerio de Ciencia,
Innovaci{\'o}n y Universidades (ref. BEAGAL 18/00148) and cofinanced by the
Universitat de les Illes Balears. RT is supported by the Spanish Ministerio de
Ciencia, Innovaci{\'o}n y Universidades (ref. FPU 18/00694). This research has
made use of data or software obtained from the Gravitational Wave Open Science
Center (gwosc.org), a service of LIGO Laboratory, the LIGO Scientific
Collaboration, the Virgo Collaboration, and KAGRA. LIGO Laboratory and Advanced
LIGO are funded by the United States National Science Foundation (NSF) as well
as the Science and Technology Facilities Council (STFC) of the United Kingdom,
the Max-Planck-Society (MPS), and the State of Niedersachsen/Germany for support
of the construction of Advanced LIGO and construction and operation of the
GEO600 detector. Additional support for Advanced LIGO was provided by the
Australian Research Council. Virgo is funded, through the European Gravitational
Observatory (EGO), by the French Centre National de Recherche Scientifique
(CNRS), the Italian Istituto Nazionale di Fisica Nucleare (INFN) and the Dutch
Nikhef, with contributions by institutions from Belgium, Germany, Greece,
Hungary, Ireland, Japan, Monaco, Poland, Portugal, Spain. KAGRA is supported by
Ministry of Education, Culture, Sports, Science and Technology (MEXT), Japan
Society for the Promotion of Science (JSPS) in Japan; National Research
Foundation (NRF) and Ministry of Science and ICT (MSIT) in Korea; Academia
Sinica (AS) and National Science and Technology Council (NSTC) in Taiwan. 
The authors are grateful for computational resources provided by the LIGO
Laboratory and supported by National Science Foundation Grants PHY-0757058 and
PHY-0823459. 
The authors gratefully acknowledge the computer resources at
Artemisa, funded by the European Union ERDF and Comunitat Valenciana as well as
the technical support provided by the Instituto de Física Corpuscular, IFIC
(CSIC-UV).

This paper has been assigned document numbers
\href{https://dcc.ligo.org/\dcc}{LIGO-\dcc}
and \href{\tdsurl}{\tds}.

\appendix
\section{\Fstat atoms}
\label{sec:fstat_derivation}
Here we will explain in more detail the procedure to obtain the \Fstat atoms,
and resulting detection statistics,
as implemented in \texttt{LALSuite}~\cite{LALSuite} and \texttt{PyFstat}~\cite{Keitel:2021xeq}.
The general approach, as originally developed for persistent CWs, is documented in Ref.~\cite{Prix:2015cfs}
though we also include here the modifications for transients following Ref.~\cite{Prix:2011qv}
and adjust some of the notation for convenience.

Detection statistics for (t)CW signals can be derived from the likelihood ratio
between hypotheses about a data set $x(t)$. In particular, a Gaussian noise
hypothesis $\mathcal{H}_{\textrm{G}}$ and a signal hypothesis
$\mathcal{H}_{\textrm{tS}}$ can be written as
    \begin{align}
        \mathcal{H}_{\textrm{G}}: x(t) &= n(t),\\
        \mathcal{H}_{\textrm{tS}}: x(t) &= n(t) + h(t;\Dop, \Amp,\TP).
        \label{eq:hypotheses}
    \end{align}
We have written this for a single frequency-evolution template $\Dop$,
but it can be generalized by iterating over a template bank.

The likelihood ratio between the two hypotheses can then be analytically
maximized \cite{PhysRevD.58.063001, Prix:2009tq} over the amplitude parameters
$\Amp$, yielding the \Fstat.
As introduced in Eq.\,\eqref{eq:fstat}, it depends on different combinations of
the start times $t^0$ and duration parameters $\tau$.
The implementation consists of computing
per-SFT quantities, the ``\Fstat atoms''.
When weighted with an appropriate window and summed up,
these give the building blocks for $2\mathcal{F}(t^0,\tau)$.

In particular, following Ref.~\cite{Prix:2011qv},
$2\mathcal{F}(t^0,\tau)$ is computed from the antenna-pattern matrix
and projections of the data onto the basis of the signal model
from Eq.\,\eqref{eq:projections}, both depending on
a window function $\win(t^0,\tau)$.
The CW case corresponds to a rectangular window covering the full observation
time.

The antenna-pattern matrix can be
written in block-form:
\begin{equation}
    \mathcal{M}_{\mu\nu}(t^0, \tau) = \mathcal{S}^{-1}T_{\textrm{data}}
    \begin{pmatrix}
        \hat{A} & \hat{C} & 0 & 0\\
        \hat{C} & \hat{B} & 0 & 0 \\
        0 & 0 & \hat{A} & \hat{C}\\
        0 & 0 & \hat{C} & \hat{B}
    \end{pmatrix}
    \label{eq:matrix_abc}
\end{equation}
where \mbox{$T_{\textrm{data}} \equiv N_{\textrm{SFT}} T_{\textrm{SFT}}$},
\mbox{$\mathcal{S}^{-1}\equiv \frac{1}{N_{\textrm{SFT}}} \sum_{X\alpha}S_{X\alpha}^{-1}$}
and $\hat{A},\hat{B},\hat{C}$ are the
independent components of the matrix given by
\begin{equation}
\begin{aligned}
    \hat{A}(t^0, \tau) &\equiv \sum_{X\alpha} \win^2(t_\alpha;t^0,\tau)\langle (\hat{a}^X_{\alpha})^2 \rangle_t, \\
    \hat{B}(t^0, \tau) &\equiv \sum_{X\alpha} \win^2(t_\alpha;t^0,\tau) \langle (\hat{b}^X_{\alpha})^2 \rangle_t, \\
    \hat{C}(t^0, \tau) &\equiv  \sum_{X\alpha} \win^2(t_\alpha;t^0,\tau) \langle \hat{a}^X_{\alpha} \hat{b}^X_{\alpha} \rangle_t.
    \label{eq:atoms_antenna}
\end{aligned}
\end{equation}
We have used here the same notation from Sec.\,\ref{sec:search}, i.e. the $X,\alpha$
indices run over detectors and SFTs, respectively. For compactness, we have suppressed the transient
parameters on the right-hand side of Eq.\,\eqref{eq:matrix_abc}.

These components are noise-weighted (indicated by the hat) such that for a
function $z^X_{\alpha}$
\begin{equation}
    \hat{z}^X_{\alpha}(t) \equiv \sqrt{w^X_{\alpha}} z^X_{\alpha}(t), \quad
\end{equation}
with weights $w_{\alpha}^X \equiv S_{X\alpha}^{-1}/\mathcal{S}^{-1}$,
and time-averaged (indicated by the brackets) such that
\begin{equation}
    \langle z^X_{\alpha} \rangle_t \equiv \frac{1}{T_{\textrm{SFT}}} \int_{t^{\alpha}}^{t_{\alpha}+T_{\textrm{SFT}}} z^X_{\alpha}(t) dt.
\end{equation}
The definitions of the antenna-pattern functions $a(t),b(t)$ can be found in
Ref.~\cite{PhysRevD.58.063001}.
In practice, $a^X_\alpha$ and $b^X_\alpha$ are computed once per SFT at its representative timestamp
instead of computing the time averages explicitly,
and noise-weighted afterwards.

The square root of the determinant of the matrix is then (suppressing again the
$(t^0,\tau)$ dependency):
\begin{equation}
    \hat{D} \equiv \hat{A}\hat{B}-\hat{C}^2 \,.
    \label{eq:determinant}
\end{equation}
Note that we have assumed the long-wavelength limit approximation \cite{Rakhmanov_2008} which implies
that $a(t), b(t)$ are real-valued and there are no off-axis terms in Eq.~\eqref{eq:matrix_abc}.

The SFT data is normalized as
\begin{equation}
    y^X_{\alpha}(t) \equiv \frac{x^X_{\alpha}(f)}{\sqrt{\frac{1}{2} T_{\textrm{SFT}} S_{X\alpha}(f)}},
\end{equation}
and used to define the two complex quantities
\begin{equation}
\begin{aligned}
    F_{\textrm{a}, \alpha}^X \equiv \int_{t_{\alpha}}^{t_{\alpha}+T_{\textrm{SFT}}} y^X_{\alpha}(t)\hat{a}^X_{\alpha}(t)e^{-i\phi^X_{\alpha}(t)}dt,\\
    F_{\textrm{b},\alpha}^X \equiv \int_{t_{\alpha}}^{t_{\alpha}+T_{\textrm{SFT}}} y^X_{\alpha}(t)\hat{b}^X_{\alpha}(t)e^{-i\phi^X_{\alpha}(t)}dt,
\end{aligned}
\label{eq:atoms_Fab}
\end{equation} 
where $\phi_{\alpha}^X(t)$ is the phase obtained from integrating the frequency
evolution of a given signal template.
These quantities take the role of the data projections $x_\mu$ in the more abstract notation used earlier,
with the detailed translation documented in Ref.~\cite{Prix:2015cfs}.

The set of $\{ \langle (\hat{a}^X_{\alpha})^2 \rangle_t, \langle (\hat{b}^X_{\alpha})^2 \rangle_t, \langle \hat{a}^X_{\alpha}\hat{b}^X_{\alpha} \rangle_t, F_{\textrm{a},\alpha}^X, F_{\textrm{b},\alpha}^X \}$
from equations \eqref{eq:atoms_antenna} and \eqref{eq:atoms_Fab} are what we refer to as the \Fstat atoms.
More technical detail of how these are implemented in \texttt{LALSuite} can be
found in Ref.~\cite{Prix:2015cfs,demod}, based on the algorithm
by Ref.~\cite{Williams:1999nt}. 

For the CNN, we use these atoms as input. On the
other hand, to compute traditional detection statistics the codes compute
Eq.\,\eqref{eq:atoms_antenna} and Eq.\,\eqref{eq:determinant} and 
\begin{equation}
    F_{\text{\{a,b\}}}(t^0,\tau) \equiv \sum_{X\alpha} \win(t_\alpha;t^0,\tau) F^{X}_{\text{\{a,b\}},\alpha} ,
\end{equation}
i.e. the window-weighted partial sums of the atoms. Finally, from all of these
it then computes
\begin{equation}
    2\mathcal{F}(t^0,\tau) = \frac{2}{\hat{D}} [
     \hat{B} |F_{\textrm{a}}|^2 + \hat{A}|F_{\textrm{b}}|^2
     - 2 \hat{C} \mathfrak{R} (F_{\textrm{a}} F_{\textrm{b}}^{\ast})  ].
     \label{eq:2F_in_terms_of_atoms}
\end{equation}
This still depends on $(t^0,\tau)$,
through the implicit partial sums for all quantities on the right-hand side.

For a search over unknown transient parameters,
one can discretize the ranges $t^0 \in
[t^0_{\textrm{min}}, t^0_{\textrm{min}}  +\Delta t^0]$ and $\tau \in [
\tau_{\textrm{min}}, \tau_{\textrm{min}}+ \Delta \tau] $ in steps $dt^0$ and
$d\tau$. Then, \mbox{$\mathcal{F}_{mn}\equiv\mathcal{F}(t^0_m,\tau_n)$} is computed for a  $N_{t^0}\times N_{\tau}$
rectangular grid
\begin{equation}
    \begin{aligned}
        t^0_m & = t^0_{\textrm{min}} +mdt^0,\\
        \tau_n & = \tau_{\textrm{min}} +nd\tau.
    \end{aligned}
    \label{eq:transient_grid}
\end{equation}

Finally, one can e.g. maximize over $\{t^0,\tau\}$, obtaining the detection
statistic \Fstatmax, or alternatively marginalize over them, obtaining the
transient Bayes factor $\Btrans$. Complete derivations of both these detection
statistics for tCWs can be found in Ref.~\cite{Prix:2011qv}.
They still depend on the frequency evolution parameters $\Dop$, and a search
where the source parameters are not exactly known will typically consist of
setting up a grid in $\Dop$ covering the desired range.

As described in Ref.~\cite{Keitel:2018pxz},
the \texttt{pycuda} GPU implementation
in \texttt{PyFstat}
is largely equivalent to the one in \texttt{LALSuite}.
(For CPU calculations of the transient \Fstat
and other functionality,
\texttt{PyFstat} calls \texttt{LALSuite} functions
through SWIG wrappings~\cite{Wette:2020air}.)
However, when testing for this paper,
we noticed that a \texttt{LALSuite} fix that was introduced in 2018
had not been included in \texttt{PyFstat} yet:
when computing $\mathcal{F}_{mn}$ over very little data
(few SFTs from a single detector),
the antenna-pattern matrix can become ill-conditioned
for some combinations of $\lambda$ parameters,
making the determinant from Eq.~\eqref{eq:determinant} approach zero
and causing spuriously large detection statistics outliers.
\texttt{PyFstat} 2.0.0
(yet to be released as of this writing)
avoids this problem the same way as the \texttt{LALSuite} code,
by truncating to \mbox{$2\mathcal{F}_{mn}=4$}
(the Gaussian noise expectation value)
when the antenna-pattern matrix condition number exceeds
a threshold that has been fixed to $10^4$ .
Results included in Sec.~\ref{sec:results} use this version.
This issue of the \Fstat for short durations has also been described
in Ref.~\cite{PhysRevD.105.124007}, where an alternative and improved
statistic was derived for this case.

\section{The effect of gaps on traditional detection statistics}
\label{sec:gaps}
\begin{figure}
    \centering
    \includegraphics[width=\columnwidth]{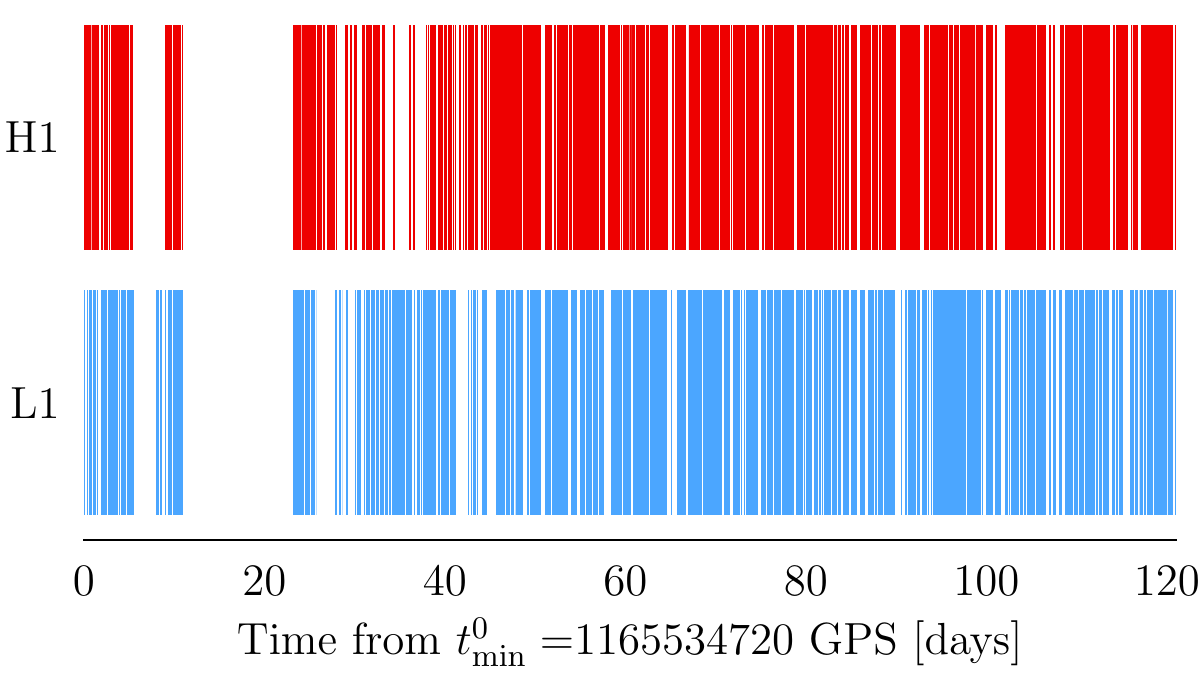}
    \caption{Time segments of the two LIGO detectors H1 and L1 corresponding to
    the data analyzed in this work. For reference the Vela 2016 glitch happened
    at $T_{\textrm{gl}}=1165577920\,$GPS, corresponding to 0.5\,days on this
    scale.}
    \label{fig:segments}
\end{figure}
\begin{figure}[t]
    \includegraphics[width=\columnwidth]{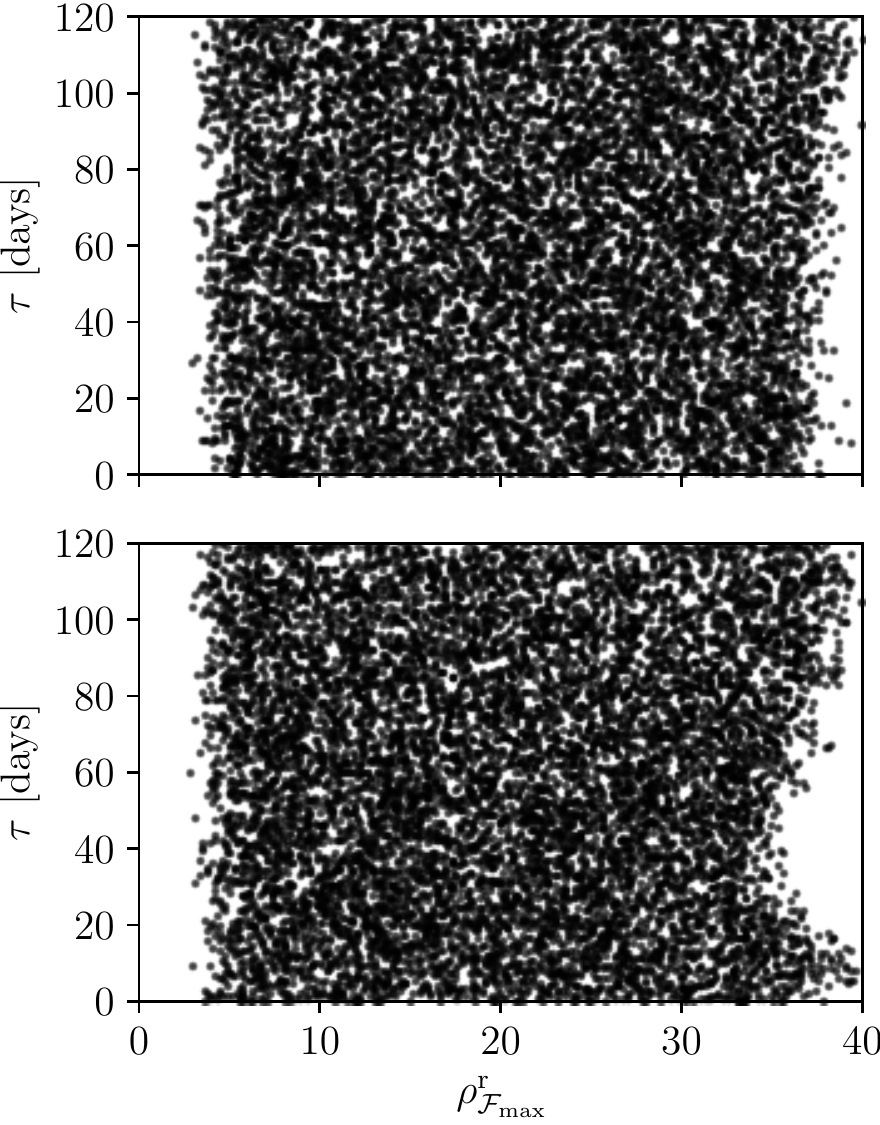}    
    \caption{Duration parameters for exponentially decaying signals and SNRs
    estimated via the traditional $2\mathcal{F}_{\max}^{\textrm{r}}$ using
    rectangular windows. In the top panel, we synthesize the statistics assuming
    no gaps in the observation time, while in the bottom panel we use the
    timestamps of O2.
    \label{fig:gaps}
    }
\end{figure}
The data we analyze in this work starts on December 11, 2016 and the final SFT
timestamp is on April 11, 2017. This time span has a notable gap of 12 days as
shown in Fig.~\ref{fig:segments}, and other smaller gaps of about 1--2 days at
most.

We have found that the CNN trained on \Fstat atoms does not show any
pathological behavior due to these gaps, though generalizing its architecture
and our training strategy to data sets with different timestamps realizations is
left to future work. On the other hand, the SFT-based \Fstat algorithm is in
general terms also robust to the presence of gaps as well, but the transient
detection statistics are somewhat affected by large gaps as the one present in
our O2 data set, depending on window choice.

First we note that if a signal falls completely inside a gap, in our testing
sets, for convenience we discard that injection and draw another set of random
signal parameters. Because in our setup $t^0$ only varies within
$T_{\textrm{gl}} \pm 0.5$\,day, this only has an effect for a small fraction of
short duration signals and for the upper limits in Sec.~\ref{sec:upper_limits},
where the shortest $\tau$ is 0.5\,days, this has no influence. 

On the other hand, if the signal falls partially into one or more gaps, behavior
is different for the two types of injections sets. In the sets where the primary
parameter is $\rho$ as used for the CNN training and the test sets in
Sec.~\ref{sec:results}, the signal amplitude is adjusted upwards to achieve the
desired $\rho$. For the upper limits injections, the $h_0$ is fixed, hence
$\rho$ is lower due to the gaps. We now concentrate on the first case.

For rectangular signals the loss will be minimal especially for long-duration
signals. When analyzing exponential-window injections with a statistic assuming
a rectangular window, however, the loss from this mismatch is worsened when the
signal falls partially in a gap.

To study this effect we generated $10^4$ synthetic samples for exponential
signals analyzed assuming rectangular windows. The parameters of the signals are
drawn from the same distributions as defined in Tab.\ref{tab:ephem}. In
Fig.~\ref{fig:gaps}, we show injected signal durations and estimated SNRs from
$\Fstatmax$ via Eq.~\eqref{eq:rho_2Fmax}. In the first panel, data without gaps
is assumed, and in the second panel we used the actual O2 timestamps, with gaps.
In the latter case there is an evident loss in estimated SNR around $\tau\sim
40$ days. This is due to the long gap of 12 days starting about 10 days after
the beginning of the analyzed data. For shorter $\tau$, most of the power of the
injected signals is concentrated before the long gap and can be recovered by a
shorter rectangular window, while there is not too much a loss of SNR from the
weaker late-time portions of the signal (spread over $3\tau$, as defined in
Eq.~\eqref{eq:winexp}) falling into the gap. On the other hand for these
durations around 40\,days, most of the power falls into the gap, hence the
noticeable loss in SNR when recovered with a mismatched rectangular window.

This effect is even more accentuated when using not only realistic timestamps
but also real data, because of the non-stationary characteristics of real noise.
While for the standard detection statistics considered here this effect cannot
be easily remedied, a properly trained CNN can alleviate the loss, as shown in
Fig.~\ref{fig:SNR_panel} for the CNN trained on a mixture of both synthetic and
real data.

\bibliography{cnnfatoms}

\end{document}